\documentclass[longauth]{aa} 
\usepackage{graphicx}
\usepackage{natbib}
\usepackage{txfonts}
\usepackage{aalongtable}
\begin{document}
   \title{Young, Massive Star Candidates Detected 
     throughout the Nuclear Star Cluster of the Milky Way}

   \authorrunning{Nishiyama \& Sch\"{o}del}
   \titlerunning{New Young, Massive Star Candidates in the MW NSC}

   \author{Shogo Nishiyama\inst{1} \and Rainer Sch\"{o}del\inst{2}
          }

   \institute{National Astronomical Observatory of Japan, 
     Mitaka, Tokyo 181-8588, Japan\\
     \email{shogo.nishiyama@nao.ac.jp}
         \and
         Instituto de Astrof\'isica de Andaluc\'ia (CSIC),
         Glorieta de la Astronom\'ia s/n, 18008 Granada, Spain\\
         \email{rainer@iaa.es}
       }

   \date{}

 
  \abstract
  {Nuclear star clusters (NSCs) are ubiquitous at the centers of
    galaxies. They show mixed stellar populations and the spectra of
    many NSCs indicate recent events of star formation.  However, it
    is impossible to resolve external NSCs in order to examine the
    relevant processes. The Milky Way NSC, on the other hand, is close enough to be
    resolved into its individual stars and presents therefore a unique
    template for NSCs in general. }
  {Young, massive stars have been found by systematic spectroscopic
    studies at projected distances $R\lesssim0.5$\,pc from the
    supermassive black hole, Sagittarius\,A* (Sgr\,A*). 
    In recent years, increasing evidence has been found for
    the presence of young, massive stars also at $R > 0.5$\,pc.
    Our goal in this work is a systematic search for
    young, massive star candidates throughout the entire region
    within $R \sim 2.5$\,pc of the black hole. 
  }
  { The main criterion for the photometric identification of 
    young, massive early-type stars 
    is the lack of CO-absorption in the spectra. 
    We used narrow-band imaging with
    the near-infrared camera ISAAC at the ESO VLT
    under excellent seeing conditions to search for young, massive stars within
    $\sim$2.5\,pc of Sgr\,A*.}
  { We have found 63 early-type star candidates at $R\lesssim2.5$\,pc,
    with an estimated erroneous identification rate of only about 20\,\%.
    Considering their $K$-band magnitudes and interstellar extinction,
    they are candidates for Wolf-Rayet stars, supergiants, or early
    O-type stars.  Of these, 31 stars are so far unknown young, massive
    star candidates, all of which lie at $R>0.5$\,pc.  The surface
    number density profile of the young, massive star candidates can
    be well fit by a single power-law ($\propto R^{- \Gamma}$), with
    $\Gamma = 1.6 \pm 0.17$ at $R < 2.5\,$pc, which is significantly
    steeper than that of the late-type giants
    that make up the bulk of the observable stars in the NSC.
    Intriguingly, this power-law is consistent with the power-law that
    describes the surface density of young, massive stars in the same
    brightness range at $R\lesssim0.5$\,pc. }
  { The finding of a significant number of newly identified early-type
    star candidates at the Galactic center suggests that young, massive
    stars can be found throughout the entire cluster
    which may require us to modify existing theories 
    for star formation at the Galactic center. 
    Follow-up studies are needed to improve the existing data and lay the
    foundations for a unified theory of star formation in the Milky Way's NSC. }

   \maketitle
%

\section{Introduction}

Nuclear star clusters (NSCs) are ubiquitous in galaxies and appear as
compact clusters at the dynamical centers of their host galaxies
\citep[e.g.,][]{Boker02HST1,Carollo98,Cote06}. 
They have luminosities in the range $10^{5}-10^{8}$\,L$_{\odot}$, 
effective radii of a few pc, and masses of the order
$10^{6}-10^{8}$\,M$_{\odot}$. They are typically 1-2 orders of
magnitude brighter and more massive than globular clusters \citep{Walcher05}, 
which places NSCs among the most massive known clusters in the Universe 
\citep[see][for a brief review on NSCs]{Boker10IAUS}. 
Star formation in NSCs appears to be a (quasi-)continuous process. 
The majority of NSCs have mixed old and young stellar populations 
and show frequently signs of star formation within the past 100\,Myr 
\citep[e.g.,][]{Walcher06}. 
NSCs show complex morphologies and frequently coexist with supermassive black holes 
\citep[SMBHs;][]{Seth06,Seth08}.

Sagittarius\,A* (Sgr\,A*), the SMBH at the center of the Milky Way (MW), 
with a mass of roughly 4 million solar masses and located 
at a distance of about 8\,kpc \citep[e.g.,][]{Ghez08PM,Gillessen09PM}, 
is surrounded by a dense and massive star cluster, 
the MW's NSC, 
which has an estimated mass of $3\pm1.5\times10^{7}$\,M$_\odot$ \citep{Launhardt02} 
and a half light radius of 3-5\,pc 
\citep[][see the latter paper for a brief review on the MW NSC]{Graham09,Schodel11ASPC}. 
The mid-infrared images from the Spitzer Space Telescope 
show how the NSC stands out as a separate structure 
at the center of the MW \citep{Stolovy06JPhCS}.

Star formation events occurred in the central parsec of the Galactic
center (GC) about $10^{8}$ and a few times $10^{6}$ years ago \citep{Krabbe95}.  
The massive, young stars formed in the most recent star formation event 
were found to be concentrated within a projected radius of $R=0.5$\,pc 
around the central SMBH, with their density increasing toward Sgr\,A*. 
About half of them appear to be located within a disk-like structure
\citep{Levin03,Paumard06,Lu09}.  
The currently favored hypothesis for their formation is that they formed
about 6 Myr ago in situ in a dense gas disk around the SMBH
\citep[e.g.,][]{Bonnell08Sc,Bartko10}.  An alternative scenario
for the presence of young, massive stars in the central parsec is
the formation of a massive cluster at more than several parsecs distance from
the SMBH, followed by its infall toward the central parsec via
dynamical friction and its subsequent dissolution
\citep[e.g.,][]{Gerhard01,Kim03}. However, this hypothesis encounters
difficulties in explaining observations such as the steep increase of
the surface density of the young stars toward Sgr\,A* and the
relatively low total mass of the young, massive stars in the central
0.5\,pc \citep[see discussion in][]{Bartko10}. Also, the apparent
absence of X-ray radiation from young, low-mass stars in the central
parsecs disfavors the cluster infall hypothesis
\citep{Nayakshin05Acc}.

Extinction toward the central parsecs of the GC is extreme,
  $A_{V}=30-50$\,mag and variable on arcsecond scales
  \citep[e.g.,][]{Scoville03,Schodel10Ext}, which means that stellar
  colors are dominated by extinction, and broad-band colors can hardly
  be used to distinguish between late- and early-type stars
  \citep[see, however, the carefully extinction corrected
  color-magnitude diagram in Fig.\,8 of][]{Schodel10Ext}.  Alternative
  methods to identify young, massive stars are spectroscopy, X-ray
  plus infrared multi-wavelength observations, search for emission
  lines, or near-infrared imaging with filters sensitive to the
  CO-absorption of late-type stars. Since spectroscopic observations
  are very intensive in observing time, they were almost exclusively
  focused toward a region within $R\sim0.5$\,pc from Sgr\,A*, 
  with only a few observed fields at $0.5 < R < 0.8$\,pc
  \citep[e.g.,][]{Do09Cusp,Bartko10}. Despite of the small area
  examined, spectroscopic observations dominate our current
  understanding of star formation at the GC with the view that young,
  massive stars can almost exclusively been found within
  $R\sim0.5$\,pc of Sgr\,A*. 

  However, evidence for the presence of
  young, massive stars in the NSC at distances $R>0.5$\,pc from Sgr\,A*
  has been steadily increasing in the past years.
 \citet{Geballe06IRS8} analyzed the spectrum of a single star,
  IRS\,8, located about 1\,pc north of Sgr\,A*, and identified it as a
  O5-O6 giant or supergiant. This source was an obvious target for
  dedicated spectroscopic observations 
  because it is surrounded by a bow-shock
  and thus stands out from the field.
  \citet{Mauerhan10X} combined X-ray and near-infrared
  observations to identify a few dozen Wolf-Rayet stars and O giants
  or supergiants throughout the central region (at $R<100$\,pc) of the MW. These
  observations are probably particularly sensitive to massive
  binaries, where the X-ray emission is produced in colliding
  winds. \citet{Wang10HST}, \citet{Dong11HST}, and
  \citet{Mauerhan10IRW} used Pa\,$\alpha$ narrow-band imaging with HST to
  find young, massive stars throughout 
  within a few tens of pc of Sgr\,A*.
  This method picks out most likely massive stars with strong winds, 
  but it runs into problems in the central parsecs near
  Sgr\,A* because of the bright emission from the interstellar medium
  in the so-called mini-spiral \citep{Ekers83,Lo83Nat} and the
  relatively low angular resolution that they used
  ($FWHM\approx0.2"$). An efficient way of distinguishing between
  young, massive stars and old, late-type stars in imaging data is 
  near-infrared imaging that makes use of the CO-absorption at
  $\lambda\approx2.3\,\mu$m in the atmospheres of the cool stars
  \citep{Genzel03Cusp,Buchholz09}. Young, massive stars are identified
  in this method through the absence of CO-absorption combined with
  their high luminosity. \citet{Buchholz09} used this method, combined
  with adaptive optics (AO) imaging and found evidence for a
  population of young, massive stars at $R>0.5$\,pc from the central
  SMBH. However, the use of AO with the correspondingly small
  instrumental field-of-view (FOV) restricted them to a relatively
  small region at $R \leq 20''/0.8\,$pc from Sgr\,A*.

  Here, we present the results of a systematic
  search for young, massive stars within $R\approx2.5$\,pc of Sgr\,A*
  via imaging data from the VLT near-infrared camera
  ISAAC\footnote{Based on observations collected at the European
    Organisation for Astronomical Research in the Southern Hemisphere,
    Chile under programme 081.B-0099.}. Our method is narrow-band
  seeing-limited imaging under good seeing conditions, using the
  CO-band absorption of late-type stars to distinguish between
  young, massive, early-type stars 
  and late-type giants in the central $\sim$$6\times6$\,pc.
  Recurrence to seeing-limited observations
  allow us to probe a significantly larger FOV than what would be
  possible in comparable time with AO imaging.  Our work has some
  overlap with, but is also complementary to the above-mentioned work.


\section{Observations and Data Reduction}
\label{sec:Obs}

Imaging observations were acquired through various narrow-band filters 
with ISAAC at the ESO VLT \citep{Moorwood98Msn}. 
The FOV is $2\farcm5 \times 2\farcm5$ with a pixel scale of 0\farcs15.
Details of the observations are listed in Table\,\ref{Tab:Obs}. 
The angular resolution of observations with ISAAC are limited by seeing, 
which is of special importance here because the GC is an extremely crowded field. 
Note that the DIMM seeing measurements listed in Tab.\,\ref{Tab:Obs}
present a pessimistic over-estimation of the actual seeing values. 
On the one hand, according to the Roddier formula, 
the seeing FWHM$\propto\lambda^{-0.2}$, hence visual seeing of $1.0"$ translates
to about $0.8"$ at $2.09\,\mu$m. 
On the other hand, the DIMM measurements on Paranal are influenced 
by a turbulent layer close to the ground that hardly affects the telescopes themselves \citep{Sarazin08Msn}. 
The average point-source FWHM in all our images is therefore only about $0.5"-0.6"$.

Most of the data reduction process was standard, with sky subtraction,
bad pixel correction, and flat-fielding. The sky observations were
acquired with identical settings as the target observations, but on a
dark cloud  about $713''$ west and $400''$ north of Sgr\,A*.

\begin{table*}[htb]
\caption{Summary of observations. \label{Tab:Obs}} 
\begin{tabular}{llllllll}
\hline
\hline
UTC Date & Filter ID & $\lambda_{c}$\tablefootmark{a} &
$\Delta\lambda$\tablefootmark{b} &  DIT\tablefootmark{c}&
NDIT\tablefootmark{d} & N\tablefootmark{e} & Seeing\tablefootmark{f}\\
   & & $\mu$m & $\mu$m & s & & & arcsec\\
\hline
2008 May 19 & NB1.71 & $1.71$ & 0.026 & 5.0 & 15 & 5 & $0.6-1.1$\\
2008 Sep 17 & NB1.71 & $1.71$ & 0.026 & 5.0 & 15 & 5 & $0.9$ \\
2008 Jul 18 & NB2.09 & $2.09$ & 0.02 & 5.0 & 14 & 5 & $0.6-0.8$ \\
2008 May 25 & NB2.25 & $2.25$ & 0.03 & 5.0 & 15 & 5 & $0.9-1.1$ \\
2008 Sep 17 & NB2.25 & $2.25$ & 0.03 & 5.0 & 15 & 5 & $0.7$ \\
2008 May 25 & NB2.34 & $2.34$ & 0.03 & 5.0 & 15 & 5 & $0.8$ \\
2008 Sep 17 & NB2.34 & $2.34$ & 0.03 & 5.0 & 15 & 5 & $0.9$ \\
\hline
\end{tabular}
\tablefoot{
\tablefoottext{a}{Central filter wavelength.}
\tablefoottext{b}{Filter width.}
\tablefoottext{c}{Detector integration time.}
\tablefoottext{d}{Number of DITs averaged per dither position by
  readout electronics.}
\tablefoottext{e}{Number of dither positions.}
\tablefoottext{f}{According to DIMM measurements.}
}
\end{table*}

A non-standard part of the image reduction process was 
taking care of detector row bias variations 
as described in the {\it Instrument Data Reduction Cookbook}, issue 80.0, 
edited by ESO Paranal Science Operations and available on the ESO web site. 
Since different rows of the detector are read out at different times 
when applying double correlated read, 
different parts of the detector sample 
different parts of the non-linear integration ramp. 
This can lead to important biases for short integration times and bright sources. 
The bias variations depend on integration time and detector illumination, 
and are not constant in time. 
In particular, the bias depends on the brightness and distribution of sources over the FOV 
and is therefore different for each pointing. 
The bias variations are non-uniform across the detector 
and cannot be modeled by a smoothly varying function. 
Since the upper and lower halves of the detector are read out in parallel 
there is a visible jump of the background around detector 
row 512 in individual exposures.

However, the bias variations are largely restricted to 
the vertical (y-)direction on the detector 
and the bias is, to 0-th order, uniform along each row. 
Therefore, the bias can, in principle, be reliably determined 
by measuring the median along each row, 
after suitably clipping the highest and lowest pixel values.  
A difficulty in this procedure is that the GC is extremely crowded 
and there are hardly any pixels on the detector 
that do not contain part of the PSF of a star.  
After some experimenting we chose to use the median of the 200 lowest pixels 
from each row to estimate the bias. 
The 200 lowest pixels usually fall on dark clouds in almost each row in each image, 
where the bias can then be determined with reasonable accuracy, 
but at the price of losing information on the absolute background emission. 
After bias correction the images were combined into a final mosaic.

\begin{figure*}[!htb]
  \includegraphics[width=\textwidth,angle=0]{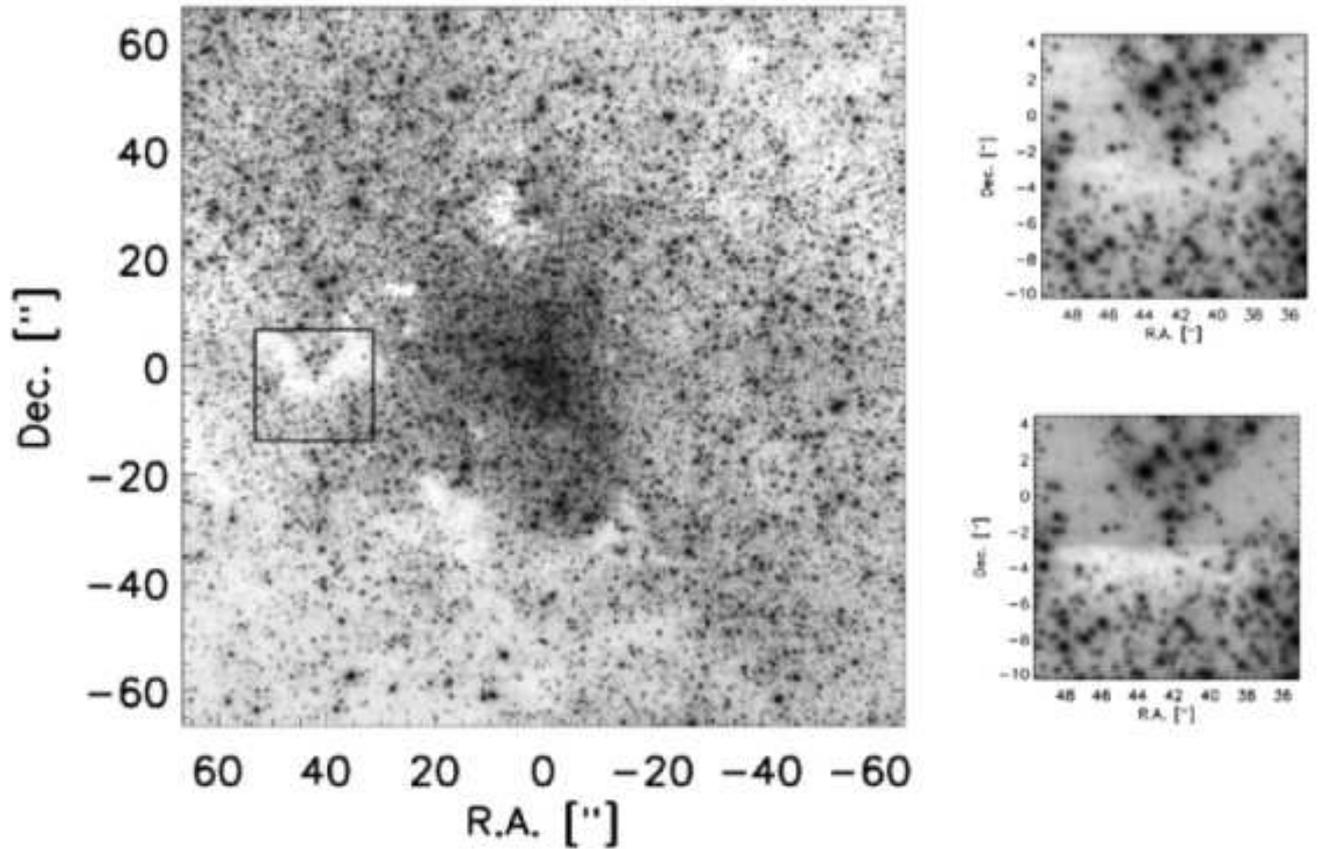}
  \caption{\label{Fig:bias} Image of the GC through the  $2.25\,\mu$m
    narrow-band filter of ISAAC (left). 
    The inserts on the right show a detail after (top)
    and before (bottom) correction of the row bias, which removes most,
    but not all, of the variable detector background.}
\end{figure*}

\begin{figure*}
  \centering
  \includegraphics[width=0.9\textwidth]{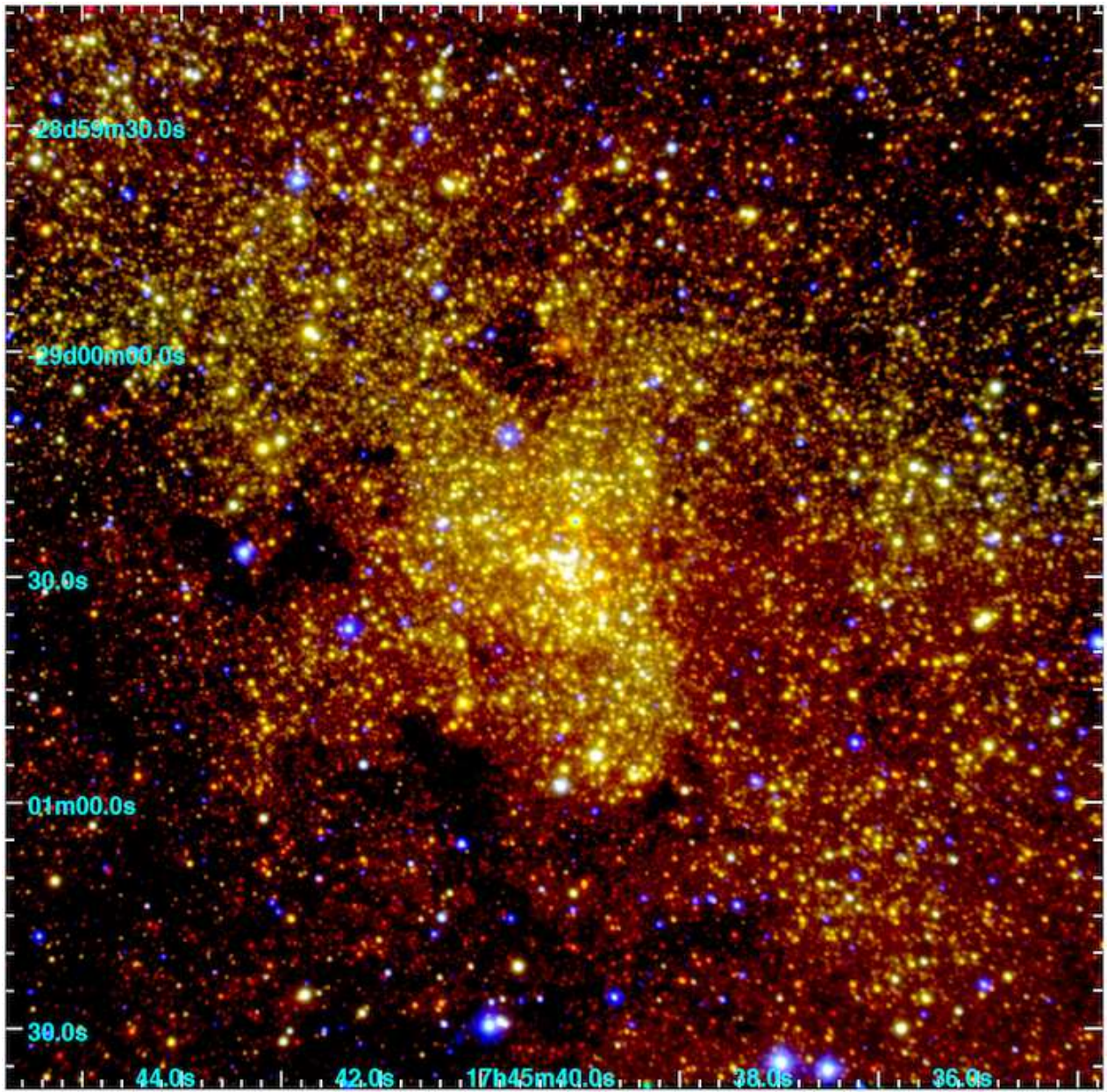}
  \caption{
    Near-infrared three-color ($1.19\,\mu$m, $1.71\,\mu$m, and $2.25\,\mu$m) 
    composite image of the central $\sim 150 \arcsec$ region
    of our Galaxy.
    East is to the left, and north is up.
  }
  \label{fig:3colorImage}
\end{figure*}

The $2.25\,\mu$m narrow-band image is shown in Fig.\,\ref{Fig:bias}, 
along with zooms onto a detail in the final mosaics composed 
from the bias-corrected and from the uncorrected images.  
The procedure for bias correction worked well, 
but some systematic errors remained, particularly because 
the brightness and density of stars is highly non-uniform across the images, 
which leads to over-estimation of the bias in rows 
crossing the center of the cluster and under-estimation
of the bias in rows that cross dark patches of strong extinction. 
Since the final mosaics were composed of dithered images,
the systematic errors of the bias correction were somewhat smeared
across the final mosaics. 
Since we used PSF fitting for photometry, with a simultaneous 
estimation of the background, 
the impact of a variable background on photometry is somewhat mitigated. 
Nevertheless, as discussed below, we could observe color trends in our data, 
mainly along the y-direction, with a jump near the middle of the images.  
We believe that these color trends are related to the uncertainties of
the bias correction.
A near-infrared three-color ($1.19\,\mu$m, $1.71\,\mu$m, and $2.25\,\mu$m) 
composite image of the central $\sim 150 \arcsec$ region
is shown in Fig. \ref{fig:3colorImage}.


\section{Data Analysis}
\label{sec:Analysis}

\subsection{Photometry}
\label{subsec:Phot}

We performed source detection and point-spread function (PSF) fitting
with the IRAF (Imaging Reduction and Analysis Facility)\footnote{ IRAF
  is distributed by the National Optical Astronomy Observatory, which
  is operated by the Association of Universities for Research in
  Astronomy, Inc., under cooperative agreement with the National
  Science Foundation.}  and DAOPHOT packages \citep{Stetson87}.  We
used the {\it daofind} task to identify point sources, and the sources
were then input for PSF fitting photometry to the {\it allstar} task.
The PSF model was constructed using a ``Penny1"
function\footnote{ A Gaussian core with Lorentzian wings.}.  From 50
to 100 sources were used to construct the variable PSF for each
image.  The PSF model was linearly variable over the image with
terms proportional to pixel offsets in the x- and y-directions.

The photometric calibration was carried out using a photometric
catalog of point sources \citep{Schodel10Ext}.  In the catalog, $H$-,
$K_S$-, and $L'$-band magnitudes of point sources in the central $40
\arcsec \times 40 \arcsec$ region are listed.  The PSF-fitting
magnitudes for the NB1.71 filter were calibrated with the $H$-band
magnitudes, and those in the NB2.09, NB2.25, NB2.34 bands were
calibrated with the $K_S$ magnitudes. The small bias introduced
  by this procedure among the NB2.09, NB2.25, and NB2.34 bands could
  be neglected for our purpose because we are primarily interested in
  differential photometry, where overall offsets are of little
  importance.  In the following, we search for early-type stars 
by using the magnitude difference between the NB2.25 and NB2.34 bands.

An astrometric calibration was done using 
the IRSF/SIRIUS point source catalog by \citet{Nishi06Ext}.
Positions of point sources in this catalog were calibrated 
with the 2MASS point source catalog \citep{Skrutskie062MASS},
and the positional accuracy of the IRSF/SIRIUS catalog
was estimated to be $\sim 0.1 \arcsec$.
Due to image distortion,
differences between stellar positions 
in the IRSF/SIRIUS catalog and ISAAC images are 
as large as $0.3 \arcsec$ at the corners of the images,
although the differences at the image center are within $0.1 \arcsec$.
This astrometric calibration was done in order for matching of 
our sources with other observations. 
The matches in stellar positions were calculated automatically at first,
but a confirmation of the matching was done by eye,
by plotting the position of the matched sources on the ISAAC images.

To avoid spurious  detections of point-like sources, 
the sources were used in the following analysis
only if a source was detected in {\it all} of the five observations 
of the NB2.09, NB2.25, and NB2.34 bands (see Table \ref{Tab:Obs}).

\subsection{Color correction and source selection}
\label{subsec:correction}

\begin{figure*}[h]
   \begin{minipage}{0.45\linewidth}
     \begin{center}
       \includegraphics[width=\textwidth]{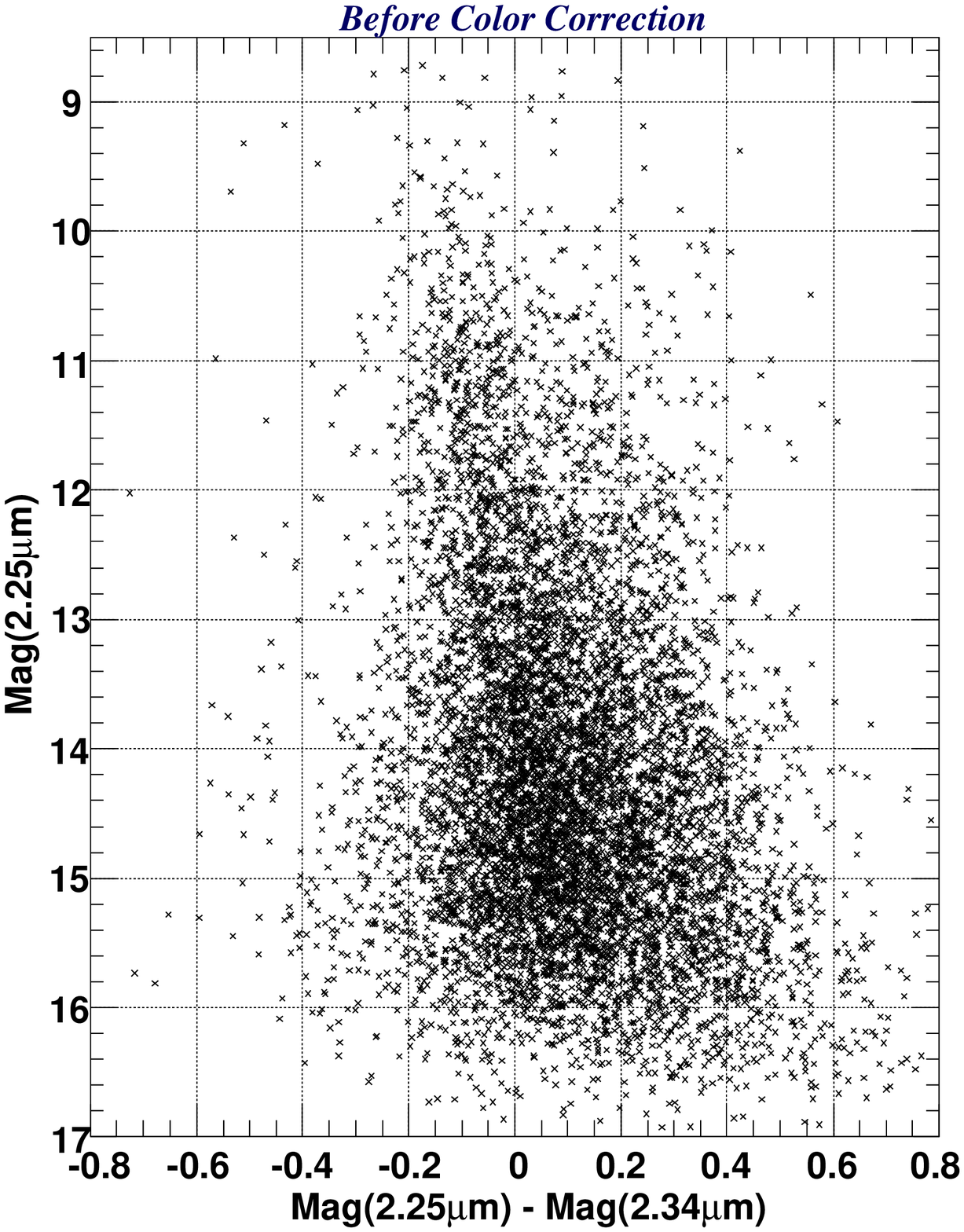}
     \end{center}
   \end{minipage}
   \begin{minipage}{0.45\linewidth}
     \begin{center}
       \includegraphics[width=\textwidth]{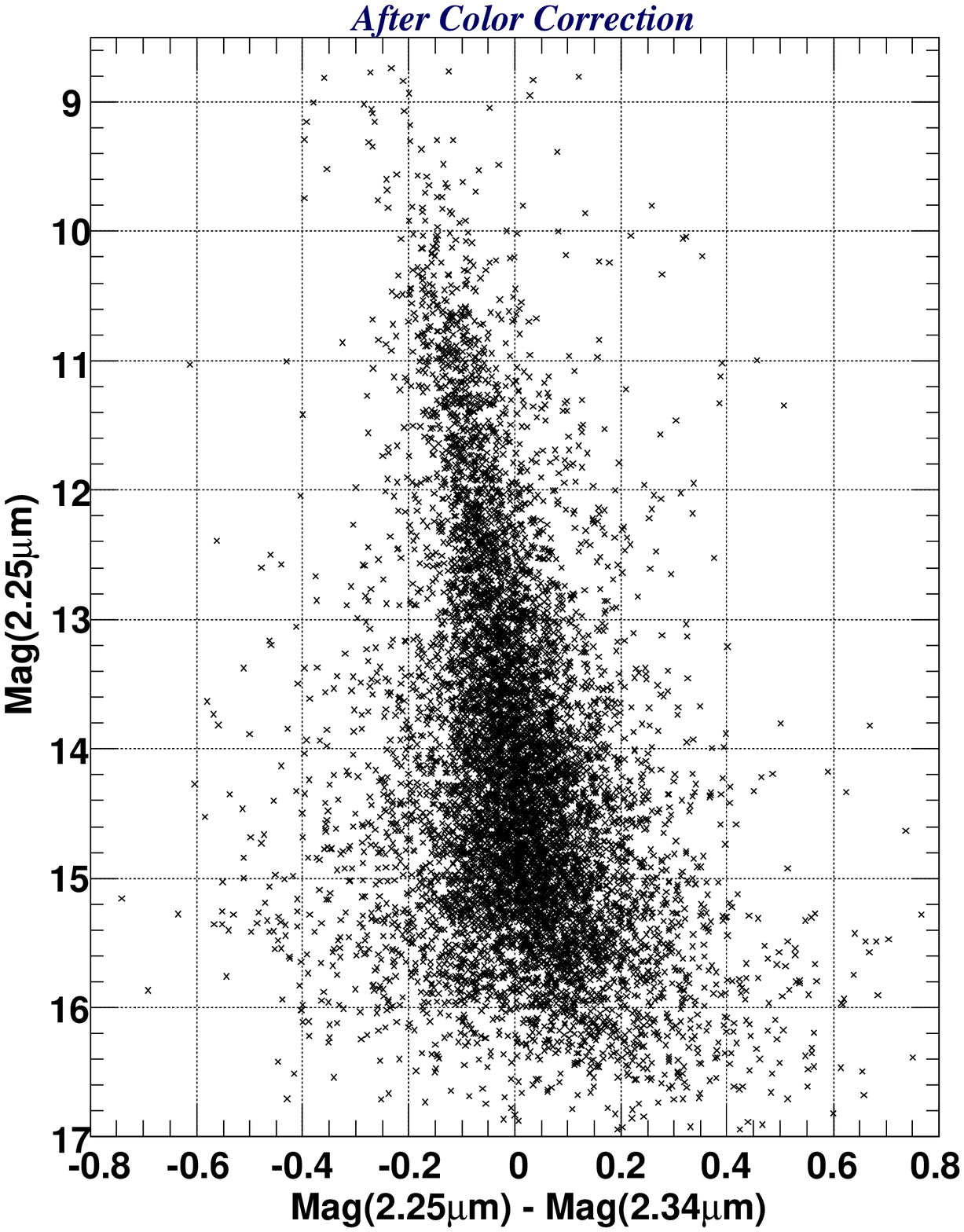}
     \end{center}
   \end{minipage}
    \caption{
      [2.25]  vs  $[2.25] - [2.34]$
      color magnitude diagrams
      before (left) and after (right) the color correction.
      See \S \ref{subsec:correction} in detail.
    }
  \label{fig:CMDColCorre}
\end{figure*}

\begin{figure}[h]
 \begin{center}
   \includegraphics[width=0.65\textwidth]{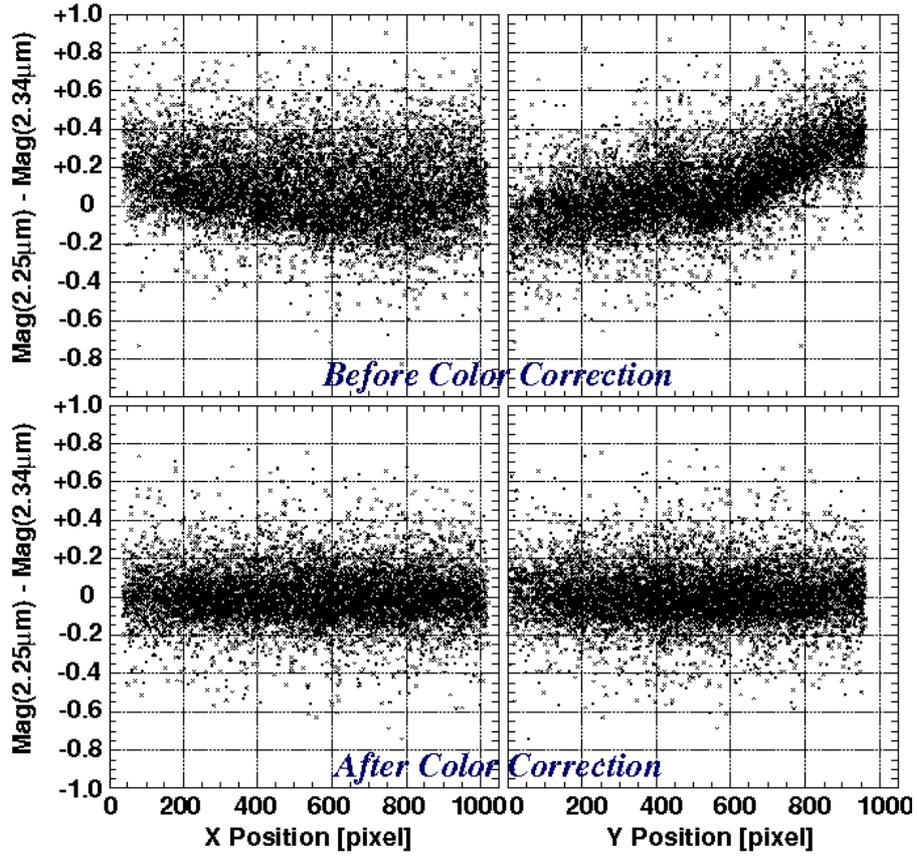}
    \caption{
      Color dependence of point sources along the x- and y-axes on the infrared array.
      We can see a clear color trend before the color correction (upper panels).
      However, after the correction (bottom panels), no clear dependence can be seen.
      See \S \ref{subsec:correction} in detail.
    }
  \label{fig:ColorPlot}
 \end{center}
\end{figure}

Although PSF fitting was done with optimized parameters
and a variable PSF,
the resulting color magnitude diagram still had a large scatter
(see the left panel in Fig. \ref{fig:CMDColCorre}).
The typical error in the $[2.25]-[2.34]$ color\footnote{ 
[$\lambda$] denotes a magnitude in a narrow-band filter
with the central wavelength of $\lambda$.}
is from 0.15 to 0.19 at $[2.25] < 13.5$,
which is larger than the expected difference of $\sim 0.1$ mag 
between early-type dwarfs and late-type giants at $\sim 2.3\,\mu$m \citep{Buchholz09}.
To investigate the cause of the large scatter,
we made $[2.25]-[2.34]$ color distribution profiles 
as functions of the image x- and y- axes (upper panels in Fig. \ref{fig:ColorPlot}).
We can see a clear color trend along the y-axis
and a discontinuity at y $\approx$ 500 pixel, 
the probable cause of which is described 
in the last paragraph of \S \ref{sec:Obs}.
The mean $[2.25]-[2.34]$ color at y $\sim 0$ is $\sim -0.1$,
but at y $\sim 950$ it is about $+0.4$.
This trend is thus identified as the principle cause of
the large scatter in the CMD.
There is also a trend along the x-axis,
but it is continuous and has a smaller amplitude.

To improve the CMD,
we carried out a color correction to reduce the influence of the color trend,
and eliminated stars with large photometric uncertainties.
For the color correction, 
we assumed that the average of the intrinsic stellar colors
is the same throughout the observed field.
This assumption is appropriate for our observed field due to
similar intrinsic colors of almost all stellar types observable around the $K$ band, 
the dominance of late-type giants in the GC region,
and the restricted wavelength range of our observations.
This assumption is also well justified in the case of GC, 
and has been used in previous work, 
e.g., \citet{Schodel07NSC,Schodel10Ext},
where this point is further discussed and illustrated.

At first, to correct the color trend along the y-axis, 
the observed field was divided into sub-fields of 1024 (x) $\times$ 20 (y) pixels.
The $[2.25]-[2.34]$ color histograms of stars in the sub-fields were constructed
and fit with Gaussian functions.
As described above, the intrinsic colors of the histograms' peaks
are expected to be almost the same,
and the difference between the central wavelengths
of the NB2.25 and NB2.34 filters is only $0.09\,\mu$m.
Preliminary color-shifts for each sub-field were therefore calculated
by requiring that the peak of each Gaussian had a color of $[2.25]-[2.34] = 0$.
After the color-shifts along the y-axis we carried out a similar procedure along the x-axis,
by dividing the observed field into sub-fields of 20 (x) $\times$ 1024 (y) pixels.
The $[2.25]-[2.34]$ color distribution profiles after the color correction
are shown in Fig. \ref{fig:ColorPlot}, bottom panels.
{\it After correction} there appears to be no significant color trend along
either the x- or the  y- axis.  The CMD after the color correction is
shown in the right panel in Fig. \ref{fig:CMDColCorre}.

Next, stars with large photometric uncertainties
were eliminated from our source list.
The stellar density is extreme in the GC, 
which means that a large fraction of sources will indeed be affected 
by systematic photometric errors due to source confusion. 
Comparing two images taken under different seeing conditions is 
an efficient way of identifying and excluding those sources from our sample.
Since two independent sets of observations existed 
for the NB2.25 and NB2.34 bands (see Table \ref{Tab:Obs}),
we could estimate the photometric uncertainties of the stars
from the differences between these two measurements.
We calculated the RMS of the photometric uncertainties 
in bins of one magnitude width,
and eliminated stars with uncertainties larger than the RMS.
Through this procedure, 3063 sources out of 8014 were excluded from our source list.
The final CMD with 4951 sources is shown in Fig. \ref{fig:CMDpastData}.
The typical combined uncertainties of the $[2.25]-[2.34]$ color
(squared sum of the photometric uncertainties in both bands)
now range from 0.06 to 0.08 at $[2.25] < 13.5$,
more than a factor of two smaller than those without
the color correction and the elimination of 
the sources with large photometric uncertainties.

\subsection{Source Classification}

\begin{figure}[h]
 \begin{center}
  \includegraphics[width=0.9\textwidth]{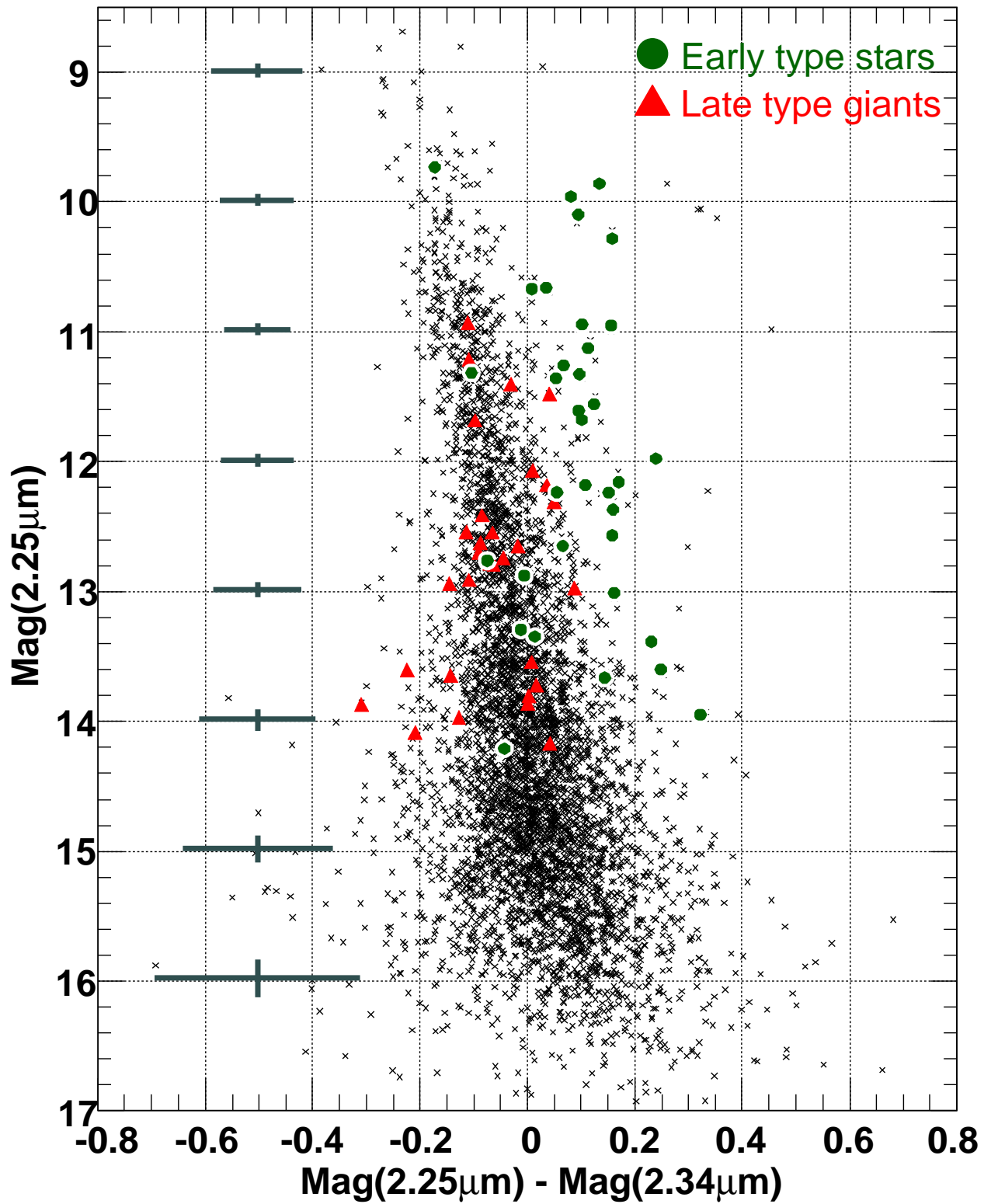}
    \caption{
      [2.25] vs [2.25] - [2.34] color magnitude diagram 
      after the color correction and 
      the elimination of sources with large photometric uncertainties.
      Known early-type stars and late-type giants are
      marked by green circles and red triangles, respectively.
    }
  \label{fig:CMDpastData}
 \end{center}
\end{figure}

The final CMD has sufficient photometric accuracy to differentiate
early-type stars from late-type giants.  In
Fig. \ref{fig:CMDpastData}, we show the $[2.25]-[2.34]$ CMD, in
  which spectroscopically confirmed late-type giants \citep{Maness07}
  and early-type stars \citep[][Table B.1]{Paumard06} are marked by
  green circles and red squares, respectively. The majority of sources
  in our final sample lie on a sequence from $([2.25]-[2.34], [2.25])
  \sim (-0.2, 9.5)$ to $(0.1, 15.5)$. This coincides with the location
  of most of the spectroscopically identified giants, indicating that
  the sequence is the red giant branch (RGB) on this CMD. On the
  other hand, at $[2.25] \la 12.5$, most of the known early-type stars
  (green circles) are located to the right of the RGB, and only two of
  a total of 23 early-type stars lie on the RGB.  Red color in $[2.25]
  - [2.34]$ means no significant CO absorption at 2.34 $\mu$m,
  thus suggesting an early spectral type.
Hence Fig. \ref{fig:CMDpastData} reveals that
early-type stars can be identified reliably 
in our sample at $[2.25] \la 12.5$.
From the location of the spectroscopically identified stars, we
estimate  a probability of false identifications on the order of 10 \%
in this region of the CMD.

\begin{figure}[h]
 \begin{center}
   \includegraphics[width=0.9\textwidth]{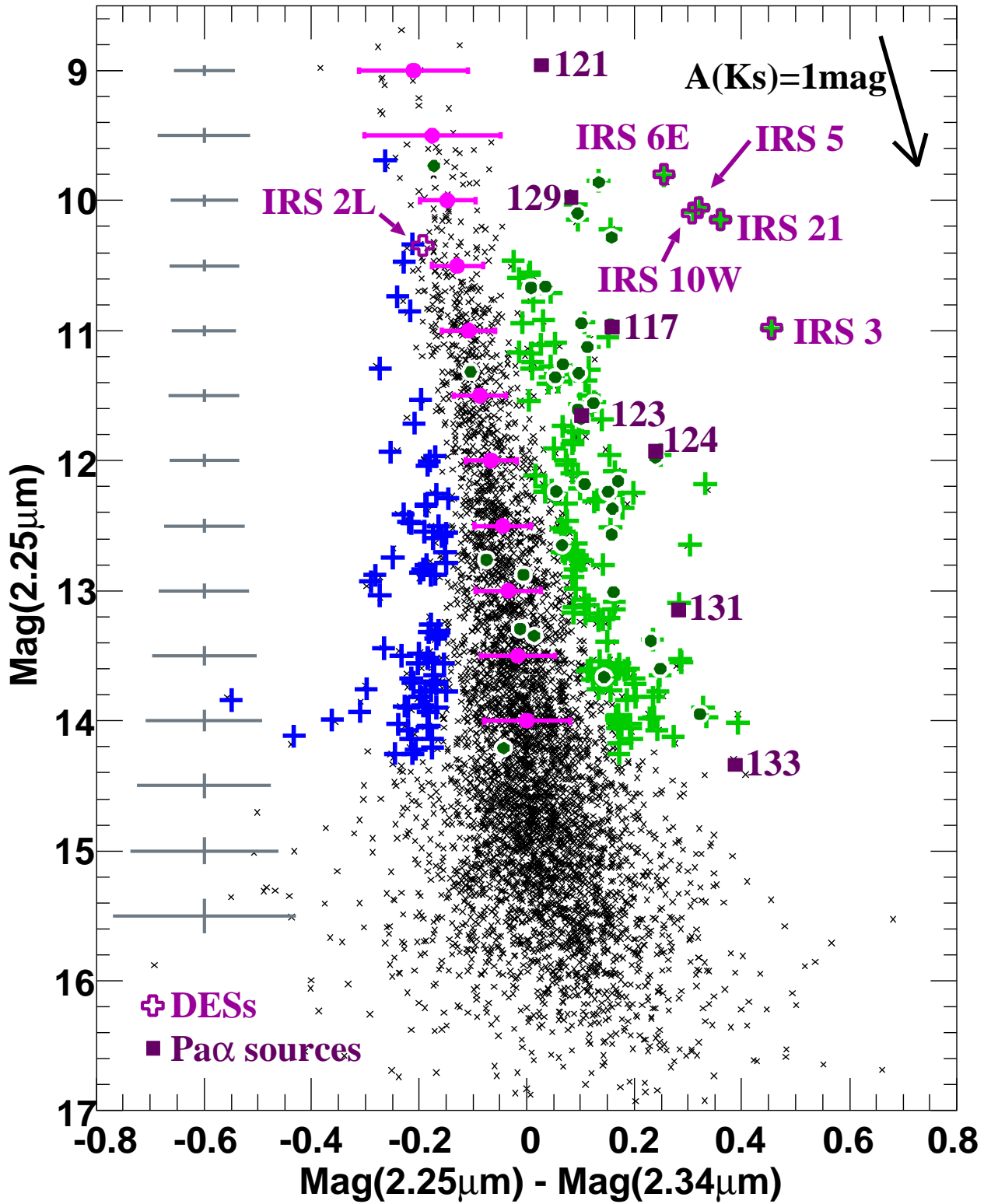}
    \caption{
      [2.25] vs [2.25] - [2.34] color magnitude diagram.
      Red color (positive value) in [2.25] - [2.34] means 
      no significant  CO absorption at $2.34\,\mu$m
      which, together with the brightness,  
      is an indicator for massive, early-type stars.
      Typical uncertainties
      are represented by grey crosses at the left side.
      Stars with photometric uncertainties larger than these typical ones
      are not included in this diagram.
      The reddening vector with a slope of $A_{\lambda} \propto \lambda^{-2.0}$
      \citep{Nishi06Ext}
      is shown at the upper right in the diagram.
      Dark green circles represent spectroscopically identified early-type stars.
      The means and sigmas of the RGB in the $[2.25] - [2.34]$ color
      within bins of one magnitude width are shown by pink circles and bars, respectively.
      We define stars more than 2-$\sigma$
      redder than the red giants as early-type star candidates (light green crosses).
      Blue crosses, which are stars more than 2-$\sigma$ {\it bluer} than the RGB,
      are used to estimate the false detection likelihood
      of the early-type star candidates.
    }
  \label{fig:CMDSelectHot}
 \end{center}
\end{figure}

To automatically identify early-type stars
and to determine the magnitude limit of our procedure,
we proceeded as follows.
We constructed $[2.25] - [2.34]$ color histograms 
for stars within one magnitude wide bins in $[2.25]$,
and fit the histograms with Gaussian functions.
The mean and standard deviation, $\sigma$, of the best fit for
each bin are shown by pink circles and pink bars in
Fig. \ref{fig:CMDSelectHot}, respectively.  The mean and $\sigma$
correspond to the mean color and width of the RGB in each
magnitude bin.  We defined stars more than 2 $\sigma$ redder than
(i.e., to the right of) the mean RGB color as early-type star
candidates (light green crosses in Fig. \ref{fig:CMDSelectHot}).

\begin{figure}[h]
  \begin{center}
   \includegraphics[width=0.7\textwidth]{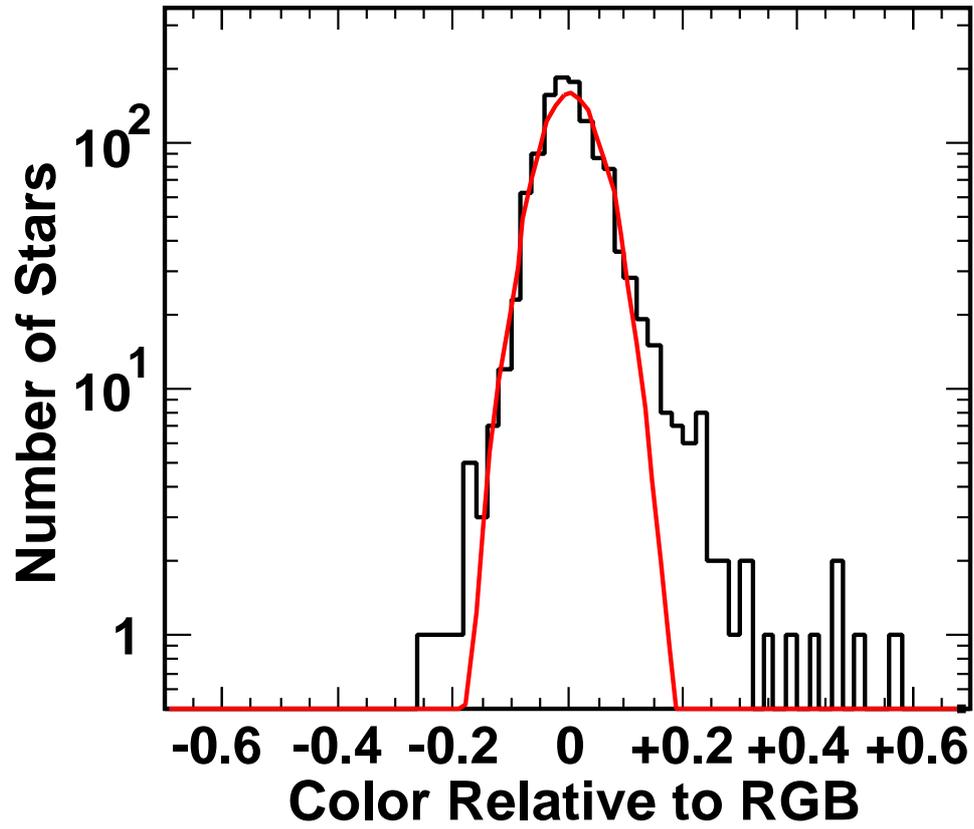}
    \caption{
      Histogram of stellar colors relative to the RGB mean
      (pink circles in Fig. \ref {fig:CMDSelectHot}).
      Only stars with $[2.25] < 12.25$ mag are used for this histogram.
      A Gaussian fit is shown by a red line.
      An excess is clearly seen on the right (redder) side of the RGB, 
      indicating the existence of early-type stars in this region.
    }
  \label{fig:ColorHistAll}
 \end{center}
\end{figure}

The color histograms of the RGB are expected to follow a normal distribution.
When we construct a histogram of 
a ``relative'' $[2.25]-[2.34]$ color from the mean of the RGB,
the histogram shows a normal distribution of the RGB
{\it plus} an excess component of the early-type stars
at the red side of the histogram (Fig. \ref{fig:ColorHistAll}).
So the ``blue'' stars, which are the blue tail 
of the normal distribution of the RGB,
can be used to estimate  the contamination of the early-type star candidates 
by RGB stars because of measurement errors.
So we define stars more than 2 $\sigma$ {\it bluer} than the RGB 
as ``blue'' stars (blue crosses in Fig. \ref{fig:CMDSelectHot}).

\begin{figure}[h]
 \begin{center}
   \includegraphics[width=0.7\textwidth]{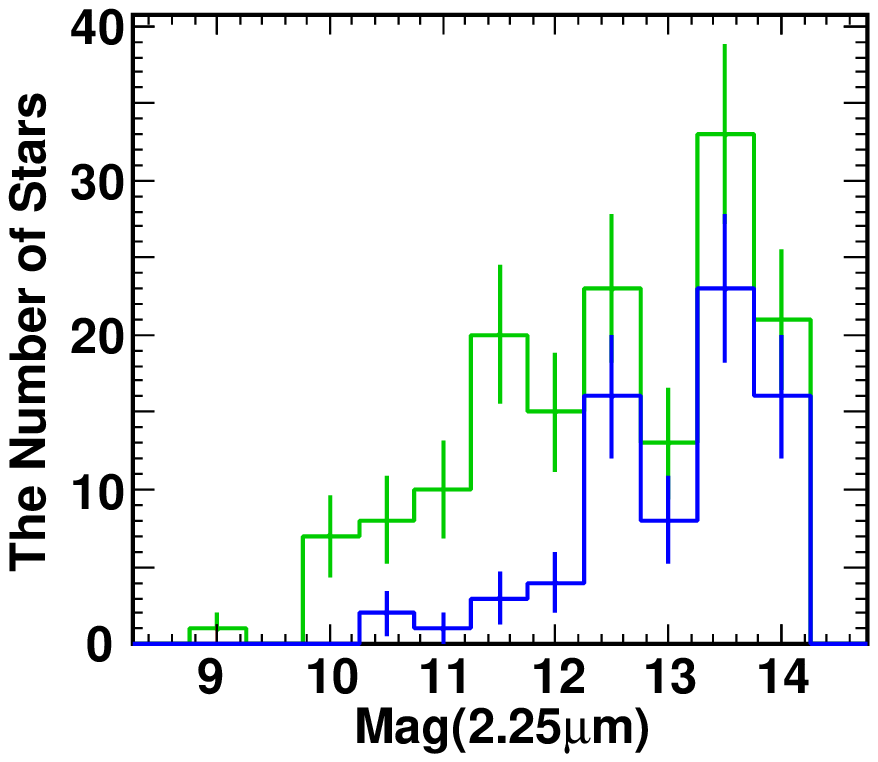}
   \includegraphics[width=0.7\textwidth]{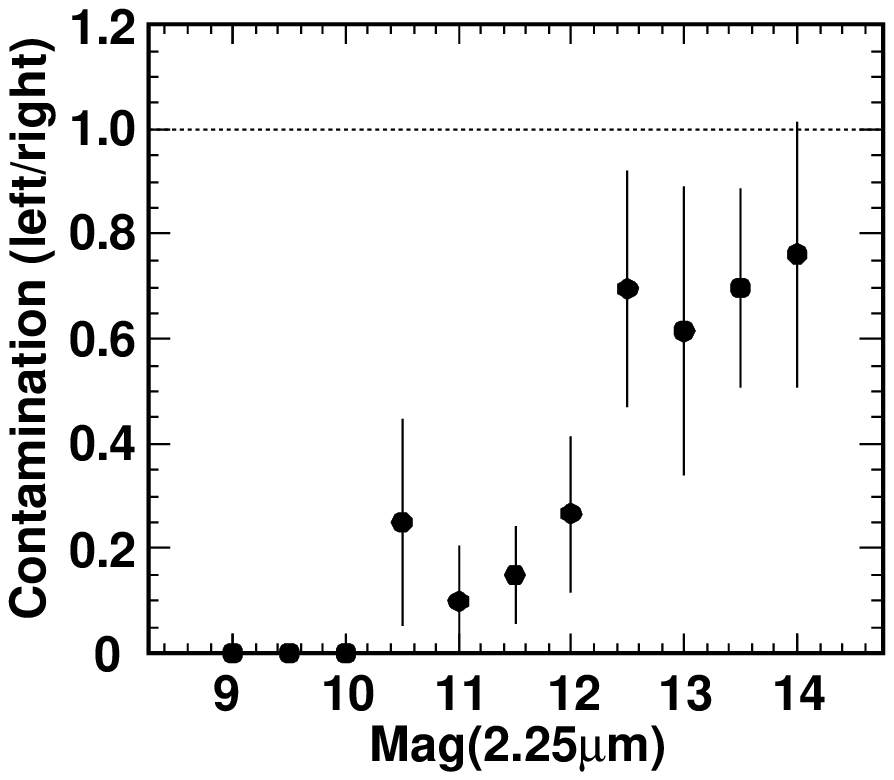}
    \caption{
      {\it Top}: Magnitude histograms for early-type star candidates
      (green)
      and stars more than 2-$\sigma$ {\it bluer} than the RGB 
      (see Fig. \ref{fig:CMDSelectHot}).
      This shows that the number of stars at the redder side 
      (earlier spectral type) of the RGB is clearly larger
      than that at the bluer side (later spectral type).
      {\it Bottom}: Ratio of the bluer stars to the early-type star candidates.
      If we assume that the bluer stars are red giants 
      with a large photometric uncertainty,
      a similar number of stars may contaminate 
      our sample of the early-type star candidates.
      So this diagram shows
      the expected ratio of such contamination for the candidates.
      The uncertainties in the top panel are given
      by the square root of the counts in each bin.
      For stars brighter than [2.25] = 12.25 mag, 
      the contamination is  low ($\lesssim 25 \%$).
    }
  \label{fig:HotSelectHists}
 \end{center}
\end{figure}

Magnitude histograms of the early-type candidates
and the ``blue'' stars are shown in Fig. \ref{fig:HotSelectHists}, top panel. 
The number of blue stars is very small at $[2.25] < 12.25$, 
but it increases at $[2.25] > 12.25$. 
The bottom panel in Fig. \ref{fig:HotSelectHists} shows
the ratio of the blue stars to the early-type candidates,
i.e., the expected contamination of the early-type candidates
by erroneous identification.
The ratios are less than $\sim$25\% at $[2.25] < 12.25$, 
and more than 60 \% at $[2.25] > 12.25$. 
Therefore we can conclude that 
the magnitude limit above which
we can differentiate the early-type star candidates from the red giants
is $[2.25] \approx 12.25$.
The stars with $[2.25] < 9.75$ are also excluded from the candidates
due to unreliable photometry caused by detector saturation.

In the magnitude range of $9.75 < [2.25] < 12.25$, 
we have found 63 early-type star candidates. 
In the same magnitude range,
20 spectroscopically confirmed early-type stars are included in the final CMD
\citep[green circles in Fig. \ref{fig:CMDpastData} and
\ref{fig:CMDSelectHot}, from ][Table B.1]{Schodel09NSC}. Of these,
only one source is mis-identified as a red giant, and the others are recognized as early-type stars in our analysis.

In the same magnitude range,
we have found 12 ``blue'' stars located at the left side of the RGB.
Assuming that the ``blue'' stars are red giants with a large photometric uncertainty,
and the number of falsely identified early-type candidates is similar,
the probability of contamination is estimated to be $12/63 = 0.19$.
However, as described below, 5 of the 63 candidates are not early-type stars,
but intrinsically very red sources.
By assuming that these two sources of uncertainty are independent, 
this increases the estimated contamination of early-type candidates to ~21\,\%. 
Note that this is a conservative estimate 
because the very red sources could be excluded by an additional color cut. 
This would, however, complicate our analysis unnecessarily. 
Also, some of these sources are probably early-type stars 
that are surrounded by bow-shocks \citep{Tanner05DES}.

\subsection{[2.25]  vs  [1.71] - [2.25] CMD }
\label{subsec:171CMD}

\begin{figure}[h]
 \begin{center}
   \includegraphics[width=0.9\textwidth]{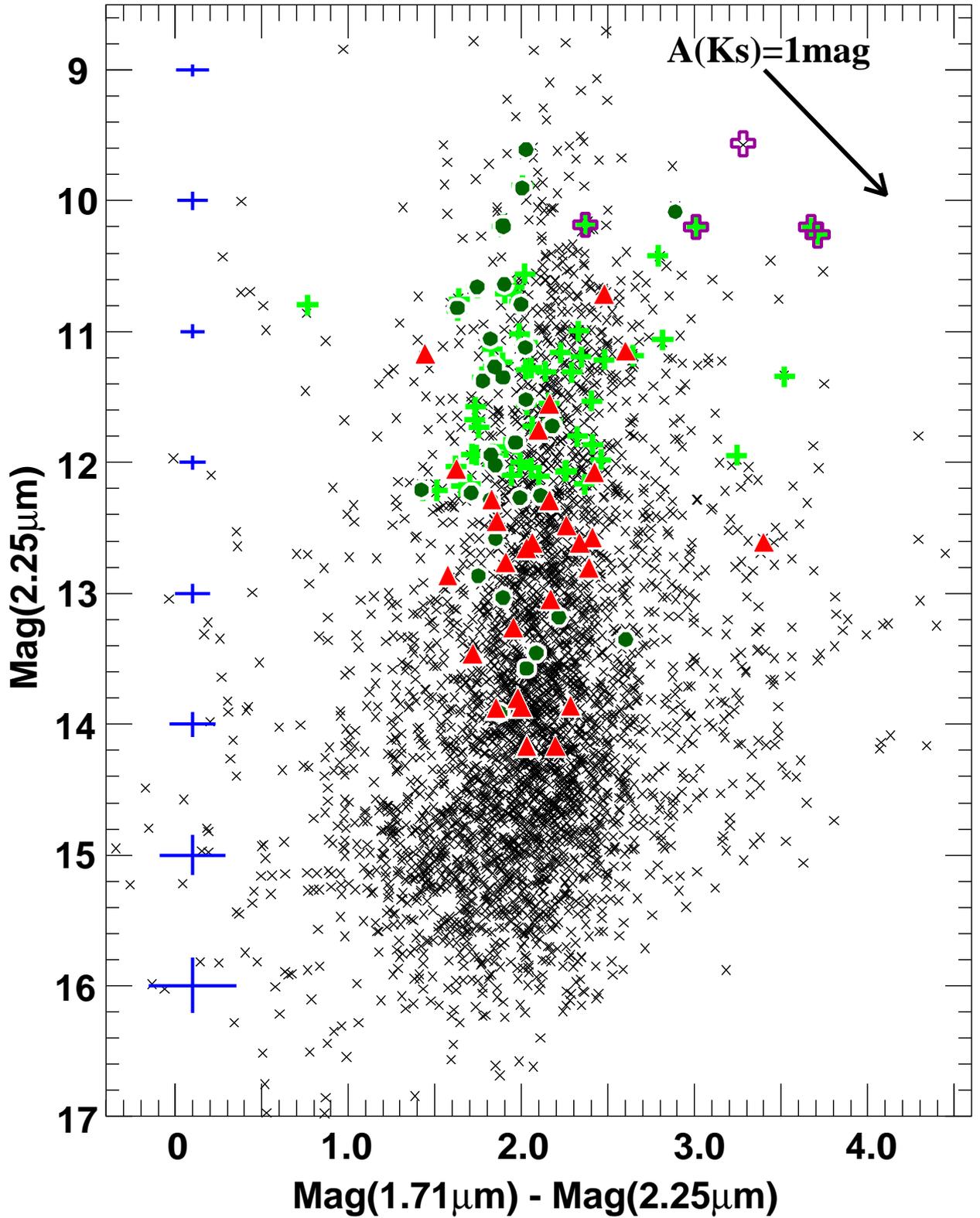}
    \caption{
      [2.25] vs [1.71] - [2.25] color magnitude diagram. 
      The early-type star candidates are overplotted by light green crosses.
      Dark green circles and red triangles
      represent spectroscopically identified early-type stars 
      and late-type giants, respectively.
      Most of the candidates have a [1.71] - [2.25] color 
      (almost equivalent to a $H - K$ color)
      of more than 1.5, suggesting that
      they suffer from strong ($A_{K} \gtrsim 2$) extinction,
      and thus they are {\it not} foreground sources.
    }
  \label{fig:CMDBH171}
 \end{center}
\end{figure}

We constructed an extinction-corrected [2.25]  vs  $[1.71] - [2.25]$ CMD 
(Fig. \ref{fig:CMDBH171})
which is expected to be very similar to a broad band $H$ vs $H - K$ CMD.
For the data sets of the NB1.71 and NB2.25 filter,
we did the same procedure of the color correction and
the elimination of the sources with large photometric uncertainties
as described in \S \ref{subsec:correction}.
Then, we carried out ``relative" extinction correction
following the procedure in \citet{Schodel10Ext};
we calculated the mean of the $[1.71] - [2.25]$ colors 
of the 10 nearest stars, $([1.71] - [2.25])_{\mathrm{mean}}$, at each position,
and the mean color excess $E_{[1.71] - [2.25]} = ([1.71] - [2.25])_{\mathrm{mean}} - 2.1$
was used for the ``relative'' extinction correction 
of the sources in the NB1.71 and NB2.25 bands.
Here we assume that most of the nearest stars are late-type red giants
which have a mean color of $[1.71] - [2.25] = 2.1$ at the GC.
More than 1-magnitude bluer ($[1.71] - [2.25] < 1.1$) or
redder ($[1.71] - [2.25] > 3.1$) sources are excluded from the nearest star list
because they are considered background or foreground stars.

By comparing both the [2.25]  vs  $[2.25] - [2.34]$ CMD 
(Fig. \ref{fig:CMDSelectHot}) and
the [2.25]  vs  $[1.71] - [2.25]$ CMD (Fig. \ref{fig:CMDBH171}),
a major benefit of the narrow-band observations with the NB2.25 and NB2.34 filters
is easily understandable.
In the [2.25]  vs  $[1.71] - [2.25]$ CMD,
it is difficult to find a difference in the distributions between
the spectroscopically confirmed early-type stars 
\citep[green circles;][]{Paumard06,Schodel09NSC}
and late-type giants \citep[red triangles;][]{Maness07}
even after our extinction correction.
This means that the accuracies of the photometry and extinction correction
are not sufficient to distinguish the early-type stars from the giants.
On the other hand, almost all of the known early-type stars are
identified as early-type star candidates in the [2.25]  vs  $[2.25] - [2.34]$ CMD 
at $9.75 < [2.25] < 12.25$ (Fig. \ref{fig:CMDSelectHot}).

Although the [2.25]  vs  [1.71] - [2.25] CMD is 
not practical for the early-type star search,
it is very useful to find foreground objects 
because the [1.71] - [2.25] color is sensitive to the interstellar extinction.
Of the early-type star candidates, only one source has 
a very blue color of $[1.71] - [2.25] \approx 0.7 $,
and it is clearly a foreground object.
However, the other sources have similarly red colors
as the spectroscopically confirmed early-type stars and giants in the GC.
They are thus likely to be located in the GC. To eliminate foreground
contamination, stars with $[1.71] - [2.25] < 1.5$ are excluded from the following
analysis.


\section{Results and Discussion}
\label{sec:Intro}

According to the definition of the early-type star candidates described above,
we have found 63 candidates in the range of $9.75 < [2.25] < 12.25$
(Table  \ref{Tab:EarlyCnad}). 
Considering their magnitude and 
the interstellar extinction of $A_{K} \sim 2 - 3$ mag in this region,
they are candidates for Wolf-Rayet stars, supergiants, or early O-type stars.

An advantage of the early-type star search with the NB2.25 and NB2.34 narrow-band filters
is an almost negligible influence of interstellar extinction.
The small difference ($0.09 \mu$m) of the central wavelengths of the two filters
introduces a very small color excess of 
$E_{[2.25]-[2.34]} = 0.07 A_{K_S}$,
according to the interstellar extinction law toward the GC
\citep[$A_{\lambda} \propto \lambda^{\alpha}$ and $\alpha \approx -2.0$, 
where $A_{\lambda}$ is the amount of extinction at the wavelength $\lambda$;][]
{Nishi06Ext,Schodel10Ext}.
In addition, the reddening vector on the $[2.25]$  vs $[2.25] - [2.34]$ CMD is
almost parallel to the RGB (Fig. \ref{fig:CMDSelectHot}).
So the objects on the CMD show 
shifts parallel to the RGB due to the interstellar extinction,
which has thus a negligible effect
on the classification of the sources.

\subsection{Comparison with other observations}

Intrinsically very red sources 
which we here call dust embedded sources (DES),
exist in the central parsec of our Galaxy
\citep[e.g.,][]{Moultaka04DES,Tanner05DES,Geballe06IRS8}.
They are included in our list of the early-type star candidates 
because there is good evidence that at least some of them are 
young, massive stars \citep{Tanner05DES}.
The position of the DESs on our CMD is shown in Fig. \ref{fig:CMDSelectHot}.
With the exception of IRS 2L, 
all of them lie far to the right in the CMD.
Several known DESs, such as IRS 1W, 9, 13,
are excluded from our final source list
because of insufficient photometric accuracy.

In the region we observed there are 18 sources 
with significant Pa\,$\alpha$ emission 
\citep[the primary Pa\,$\alpha$ emitting sources in][]{Dong11HST}.
Since most of them are concentrated within 0.5\,pc of Sgr A*,
where source confusion is severe in our images,
only seven sources are found to be matched with our final list
\citep[source ID 117, 121, 123, 124, 129, 131, 133 in][]{Dong11HST}
\footnote{\# 121 is too bright, and \# 131 and 133 are too faint 
for reliable identification of their spectral type,
so they are not included in our list of early-type star candidates}.
The sources (122, 127, 128)
are not identified as point sources due to source blending, 
one (128) is too faint for reliable detection,
and the others (38, 118, 119, 120, 125, 126, 132)
are excluded from our list
due to large photometric uncertainties.
As shown in Fig. \ref{fig:CMDSelectHot} (purple squares),
all of the seven sources in our list,
one of which is as faint as $[2.25] \sim 14$,
are distributed on the right side of the RGB.
It follows that our method can reliably identify young, massive
stars that were found via their Pa\,$\alpha$ emission.

After excluding the DESs, we find that 20 of our early-type candidates 
lie within the fields observed with
VLT/SINFONI \citep{Bartko10}\footnote{ 
Since their source list has not been published, 
we made a matching by eye with their Fig. 1.}.
Of those, 18 sources are confirmed as early-type stars
with spectroscopic observations.
This fact supports our claim that
the sources distributed on the right side of the RGB on the CMD 
are reliable candidates for early-type stars.

Spectra of six of our early-type candidates were
obtained by \citet{Blum03}, who identified four of them as
hot/young stars (IRS 3, 6E, 21,15NE) and two of them as late-type
giants (see Table \ref{Tab:EarlyCnad}). 

Twenty-eight of our early-type star candidates, after having excluded the DESs,
have been previously observed with VLT/NACO and narrow-band
filters \citep{Buchholz09}.  
Although six of those sources were
classified as late-type stars, 18 of them were listed as high
quality early-type candidates.  Two of the remaining sources could
not be classified by \citet{Buchholz09} because of a noisy (IRS 16SW)
or strongly reddened SED (ID 480), 
while a further two  sources are not listed in their catalog.

To summarize the comparison with past observations:
we have found seven Pa\,$\alpha$ sources \citep{Dong11HST} in our source list,
and all of them are identified as early-type star candidates;
20 of our candidates are included in the observed fields by \citet{Bartko10},
and 18 of them are spectroscopically classified as young, massive stars;
\citet{Blum03} observed spectroscopically six of our candidates 
and report four of them as early-type stars;
18 among 28 matches 
between our candidates and stars
from the AO imaging observations by \citet{Buchholz09} are classified
as high-quality early-type star candidates in their work,
and only six of the matches are classified as late-type stars.
The complete list of young, massive star candidates found in our
observations, along with their measured  colors, is
contained in Table \ref{Tab:EarlyCnad}.

\subsection{Spatial distribution}
\label{subsec:SpDist}

\begin{figure}[h]
 \begin{center}
   \includegraphics[width=0.9\textwidth]{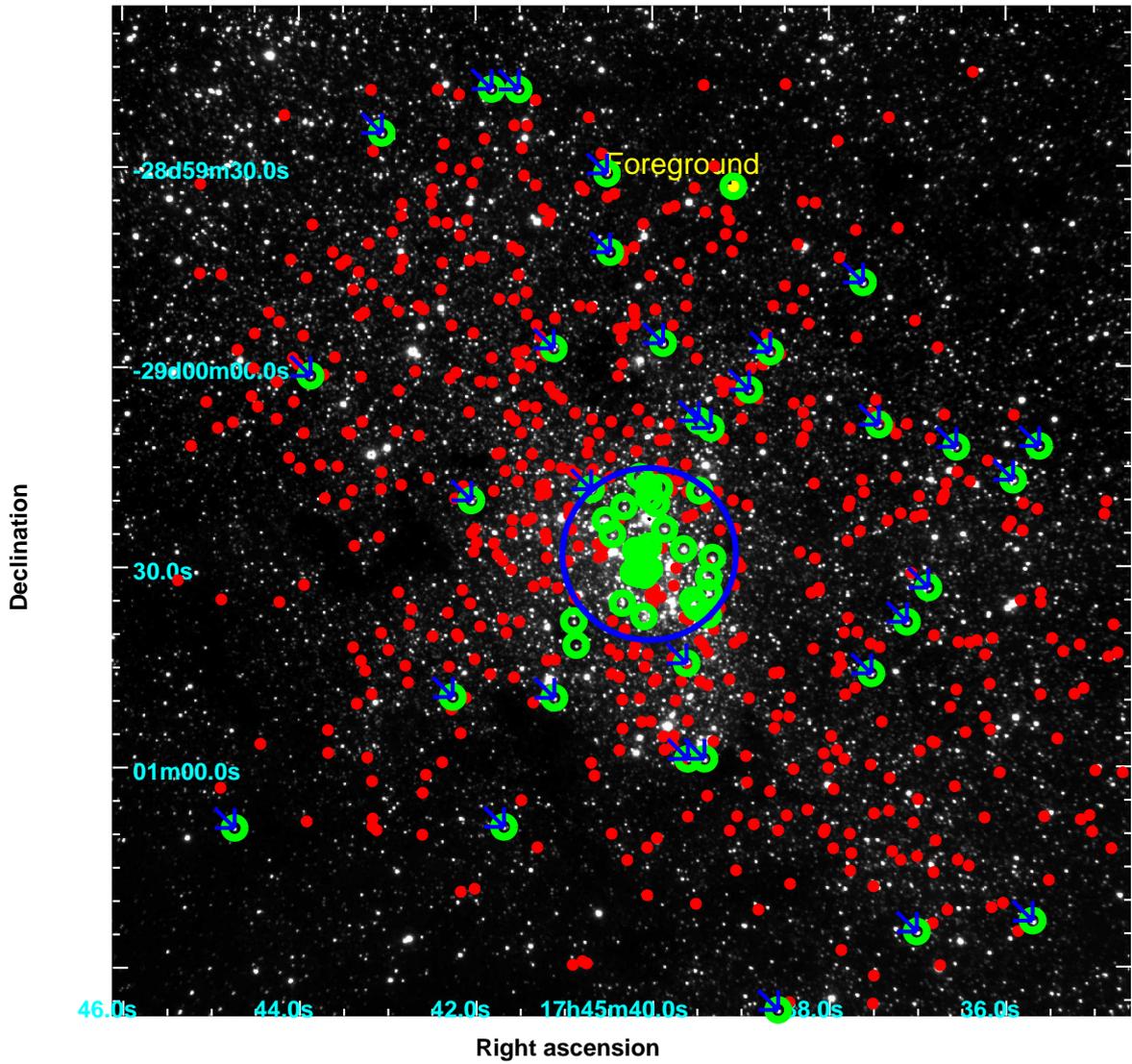}
   \caption{ Spatial distribution of the early-type star
       candidates with $9.75 < [2.25] < 12.25$ (green circles) within
       the FOV of our observations.  The candidates which have been
       unknown so far are indicated by blue arrows.  Red giants
       identified by our analysis are marked by red circles.  The
       large blue circle delimits a region within 0.5\,pc (12\farcs9) 
       in projection from Sgr A*.  There is a strong
     concentration of the early-type star candidates in the central
     0.5\,pc, as already known from previous studies.  In
       addition, we have identified about 30 new candidates outside
       the 0.5\,pc region.  Note that the density of classified
       sources decreases strongly toward the north-eastern,
       north-western , and south-eastern corners due to unreliable
       photometry in these regions, caused by strong PSF
       variability. }
  \label{fig:SpDist}
 \end{center}
\end{figure}

In Fig. \ref{fig:SpDist}, we show the NB2.25 filter image
on which we have marked
the early-type star candidates (light green circles) and 
the stars on the RGB (red dots) with $9.75 < [2.25] < 12.25$.
There is a significant concentration of the early-type candidates within a
projected distance $R_{\mathrm{Sgr A*}}$ of 0.5\,pc from Sgr A*. 
These young, massive stars are well
known from previous imaging and spectroscopic observations
\citep[e.g.,][]{Paumard06,Do09Cusp,Bartko10}.  However, more than half
(35) of our candidates are found at $R_{\mathrm{Sgr A*}} > 0.5\,$pc,
in regions that have so far hardly been examined with spectroscopic observations.
Although spectroscopic confirmation is necessary,
the large number of the candidates,
the small contamination probability of $\sim 20$ \%,
and the findings by other authors \citep[e.g.,][]{Dong11HST}
suggest the existence of a significant number of young, massive stars 
at distances $R>0.5$\,pc from Sgr A*.

The {\it intrinsic} distribution of young, massive stars will of course
differ from the observed one
because the probability of identifying a candidate at a given location
depends on factors such as extinction, crowding and local PSF.
The apparent lack of sources 
in the south eastern and north western parts of the FOV is due to
extremely high extinction in these regions
\citep[see][]{Schodel07NSC}.  Due to strong PSF distortion near the
corners and edges of the FOV of our observations, the density of
sources in our sample (both the red giants and early-type stars)
decreases significantly toward the south-eastern,
north-western,  and  north-eastern corners of the FOV.  
However, none of these difficulties affects our main finding of
a significant number of so far unknown 
candidates for young, massive stars
at projected distances between 0.5 and 3.2\,pc from Sgr\,A*.

\begin{figure}[h]
 \begin{center}
   \includegraphics[width=0.8\textwidth]{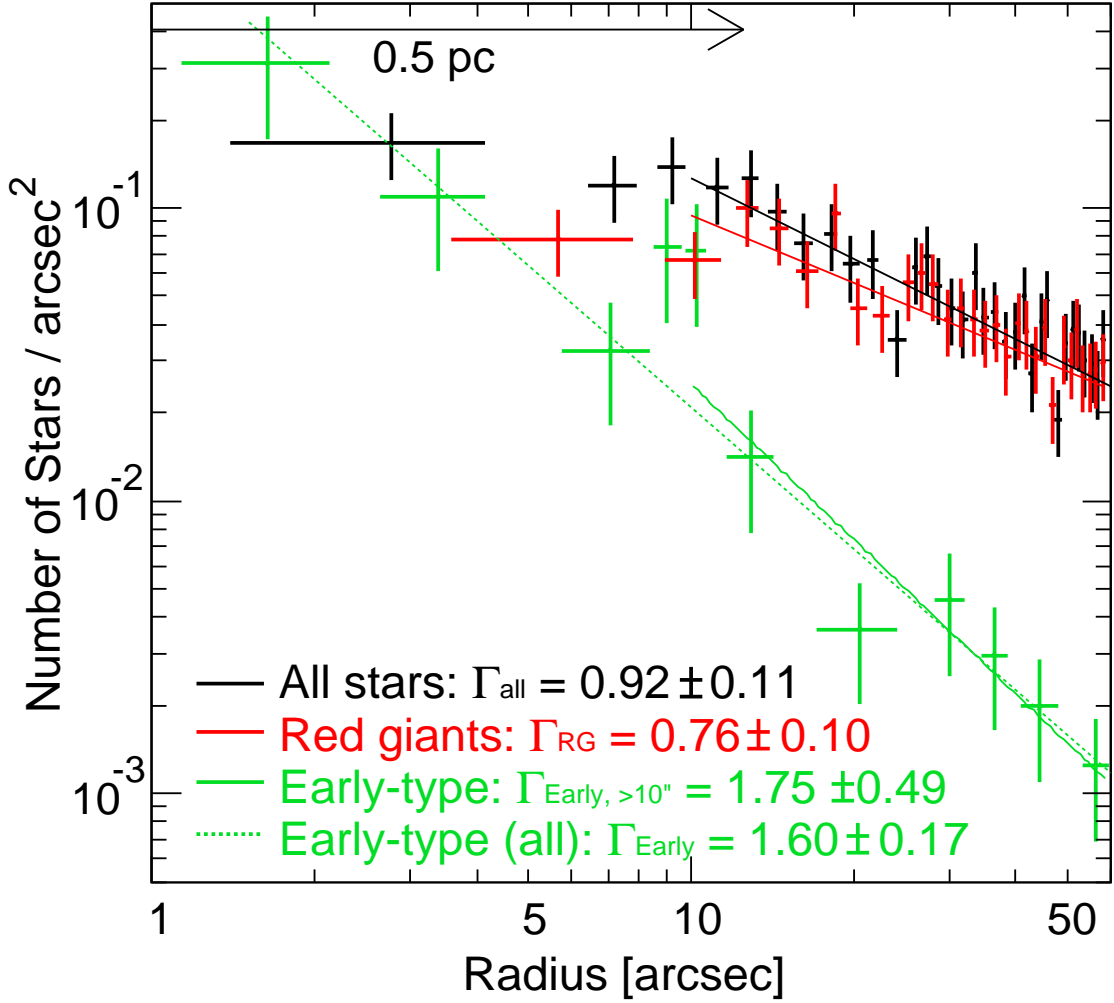}
   \caption{ Plots of the azimuthally averaged, stellar surface number
     densities as a function of projected distance from Sgr\,A* for
     all stars in our sample (black), red giants (red), and
     early-type star candidates (green), with $9.75 < [2.25] < 12.25$.
     The densities were determined in annuli of variable width,
     chosen such that each one contained a fixed number of stars (5
     stars for the early-type star candidates, and 15 stars for the
     entire sample and the red giants). Data points at $10 \arcsec
     \leq R_{\mathrm{Sgr A*}} \leq 60 \arcsec$ were used for power-law
     fits of the radial profiles for our entire sample (black
     line), red giants (red line), and early-type star candidates
     (green line).  The green dashed line represents a power-law
       fit to the densities of the early type candidates, 
       including those at $R_{\mathrm{Sgr A*}} \leq 10 \arcsec$.}
  \label{fig:radPlot}
 \end{center}
\end{figure}

Azimuthally averaged, projected stellar surface density plots as a
function of the distance to Sgr\,A* for all stars, red giants, and
early-type candidates with $9.75 < [2.25] < 12.25$ are shown in
Fig. \ref{fig:radPlot}.  This analysis is limited to the region within
$60\arcsec$ from Sgr\,A*, to extract stellar surface densities in full rings, 
and to exclude the regions with low completeness 
near the corners of our images.
The bin widths were chosen such that each bin contains a constant number of stars
(five stars for the analysis of early-type star candidates only,
and 15 stars otherwise)
in order to give each data point roughly the same statistical weight.

The surface density profiles show clearly that the spatial
distribution of young, early-type stars is different from that of old,
red giants outside the central 0.5\,pc.  We find an apparently
continuous surface density profile for the early-type star
candidates in the range $1\farcs5 \leq R_{\mathrm{Sgr A*}} \leq
60\arcsec$.  The profile was fit with a power law ($\propto R^{-
  \Gamma}$) which gives a spectral index of $\Gamma_{\mathrm{Early}} =
1.60 \pm 0.17$.  On the other hand, the profiles of the entire sample
and for the red giants show a break around $10\arcsec$ \citep[as
  known from previous work, e.g.,][]{Eckart93GC,Schodel07NSC},
so the fitting was carried out only in the range of $10\arcsec
\leq R_{\mathrm{Sgr A*}} \leq 60\arcsec$, providing power-law
indices of $\Gamma_{\mathrm{all}} = 0.92 \pm 0.11$ and
$\Gamma_{\mathrm{RG}} = 0.76 \pm 0.10$ for all stars and red giants,
respectively.  For comparison, we also fit the profile for the
early-type star candidates only for $R_{\mathrm{Sgr A*}} =
10\arcsec$ to $60\arcsec$, and obtain
$\Gamma_{\mathrm{Early,>10\arcsec}} = 1.75 \pm 0.49$.  Note that we
have {\it not} made any extinction and completeness corrections for
these plots, and thus the {\it absolute} values of $\Gamma$ may
have additional systematic errors; however, it is possible to compare
the three indices, $\Gamma_{\mathrm{all}}$, $\Gamma_{\mathrm{RG}}$,
and $\Gamma_{\mathrm{Early}}$ (or
$\Gamma_{\mathrm{Early,>10\arcsec}}$), because they suffer almost the
same interstellar extinction and have the same photometric completeness limits. 
Also, the power-law indices inferred here agree well with
previous work (see below).  The early-type star candidates show a
steep, apparently continuous radial profile, while the red giants show a
shallower profile with a break at $\sim 10\arcsec$.  The significant
difference of the profiles also suggests that not every early-type
star found in our analysis is contamination by erroneous
identification of a late-type giant.

While the power-law  indices derived here may suffer from some
  additional systematic uncertainties as discussed above,
those for the entire sample and the red giants, 
$\Gamma_{\mathrm{all}} = 0.92$ and $\Gamma_{\mathrm{RG}} = 0.76$ 
are in good agreement with previous observations.
\citet{Schodel07NSC} produced a fit of a broken power law
to the surface number density profile of the stars in the NSC,
regardless of type,
and found $\Gamma = 0.75 \pm 0.10$ outside 
a break radius at $6\farcs0 \pm 1\farcs0$.
Through narrow-band imaging observations,
\citet{Buchholz09} obtained $\Gamma = 0.86 \pm 0.06$ and  $0.70 \pm 0.09$
for all sources and late-type stars, respectively,
at $6\arcsec \leq R_{\mathrm{Sgr A*}} \leq 20\arcsec$.
Here our limiting magnitude, $[2.25] = 12.25$, is much shallower
than $K = 17.75$ \citep{Schodel07NSC} and $K = 15.5$ \citep{Buchholz09}.
So the agreement implies that incompleteness and variable extinction
do not affect our result significantly within
the photometric limit of this work.

Interestingly, the power-law indices for the early-type star candidates
are also similar to previous results at $R_{\mathrm{Sgr A*}} \la 10\arcsec$.
We construct a plane-of-the-sky projected 
surface density plot for early-type stars with $9.75 < K < 12.25$
catalogued in \citet{Bartko09GC},
and find $\Gamma_{\mathrm{Early}} = 1.47 \pm 0.42$.
 This is  consistent, within the uncertainties, with our values of 
$\Gamma_{\mathrm{Early,>10\arcsec}} = 1.75 \pm 0.49$,
and also $\Gamma_{\mathrm{Early}} = 1.60 \pm 0.17$.

\subsection{Implications for Star Formation at the Galactic Center}

In the central parsec, where the potential is dominated by the
$4\times10^{6}\,M_{\odot}$ SMBH, the densities for a molecular
cloud to be stable against tidal disruption are on the order of a few
$10^{10}\,(10^{8}, 10^{7})$ \,$n_{\rm H2}$\,cm$^{-3}$ at $r=0.1
\,(0.5, 1.0)$\,pc distance from Sgr\,A*. These extreme densities are
several orders of magnitude larger than what is found in molecular
clouds near the GC \citep[see, e.g., discussion in][]{Genzel03Cusp}. 
Standard {\it in situ} star formation is
therefore considered to be impossible in the central parsec of the
MW. Two alternative scenarios were therefore suggested to
explain the presence of young, massive stars in the central parsec:
(1) formation of the stars in a fragmenting gas disc or streamer
around the SMBH \citep[e.g.,][]{Nayakshin07Acc}; (2) formation
of a massive, dense stellar cluster at more than several parsecs distance and
its subsequent infall and dissolution \citep[e.g.,][]{Gerhard01}.

The young, massive stars at the GC were found to be concentrated
mainly within $R\sim0.5$\,pc, with the highest concentration near
Sgr\,A* and a steep fall-off of their surface density toward larger
distances. Along with some evidence for a very top-heavy initial mass
function of these stars \citep[][see however Do et al., in
preparation]{Bartko10} and with the absence of X-ray emission
from low-mass pre main sequence stars from a potential dissolved
progenitor cluster \citep{Nayakshin05Acc}, this appears to favour the
scenario of star formation in a gaseous disk around Sgr\,A*.

It has to be kept in mind, however, that our current ideas on star formation 
at the GC are influenced by the limitations of observational studies.  
Those studies have been focused largely on the region 
within a projected radius of $R\lesssim1$\,pc from Sgr\,A*. 
The main reason for this limitation is that spectroscopic searches are 
very costly and are therefore limited to small numbers of stars and/or to small areas.
Therefore it is still unclear whether a clear outer edge of 
the distribution of the early-type stars exists or not.  
Recently, further spectroscopic observations found about 10 new early-type stars 
outside the central $0.5$\,pc \citep{Bartko10}, 
and as described in the introduction, 
ever more observational results support 
the existence of young, massive stars at larger distances from the SMBH.

Here we present the results of the first systematic search, via
narrow-band imaging, for young, massive stars within a projected
distance of $R \sim 2.5$\,pc of Sgr\,A*.  Although our survey is
limited to the brightest ($K < 12.25$) stars in the field because of
the limited photometric accuracy imposed by the seeing-limited
observations in this heavily crowded field, we detect a significant
number of candidates for young, massive stars distributed throughout
the studied region, with half of them located at $R>0.5$\,pc.  The
brightness range of these candidates suggests that they are of similar
type -- and therefore age -- than the young, massive stars in the
central $0.5$\,pc.  As Fig.\,10 shows, these stars appear to be 
 randomly throughout the NSC (small number statistics
keep us from studying this aspect more quantitatively).  Intriguingly,
a single power-law, with an exponent of $\Gamma=1.60 \pm 0.17$, can
provide a good fit to the surface density of the young, massive stars
as a function of distance from Sgr\,A* throughout the studied region.
This power-law agrees within the uncertainties with the power-law that
describes the surface density of correspondingly bright young
stars within $R=0.5$\,pc of Sgr\,A*, that were identified in
spectroscopic studies \citep[e.g.,][]{Bartko09GC}.

Our finding that a single power-law can describe the surface
density of early type candidates throughout the examined region may
indicate a physical connection between the young, massive stars
inside and outside of the central 0.5\,pc.  Can these stars have
formed in the outer parts of the hypothetical gas disk?  It seems to
be difficult to form massive stars via this mechanism at $r >
0.5$\,pc. \citet{Hobbs09SFGC} suggest that the young stars at the GC
could have been formed in a high-inclination collision of two massive
gas clouds.  In their simulation, gaseous streams or filaments, 
that extend as far out as 1\,pc from Sgr\,A*, become
self-gravitating and form stars.  However, most of the stars formed
in the streamers are low mass stars ($0.1-1\,M_{\sun}$), while the
most massive stars are concentrated toward the central $0.5$\, pc
region, in contrast with our observations.  \citet{Mapelli12SFGC} also
show that in the outer part ($r > 1\,$pc) of the hypothetical
disk, the gas density is lower and thus the Jeans mass becomes about
an order of magnitude larger than in the central few tenths of a
parsec.  Hence, the resultant star formation efficiency is very low
outside a few tenths of a parsec.  Since most of the existing
theoretical work has focused on explaining the presence of young,
massive stars at $r < 0.5$\,pc, without having observational evidence
for a significant population of young, massive stars at larger
distances, they were not optimized to explain star formation at
greater distances.  Therefore, further studies are necessary to
investigate whether star formation in infalling clouds/gaseous disks
can explain the presence of young, massive stars beyond $r\sim1$\,pc
as well. Also, deeper photometric studies, similar to the one
described here, are needed to set tighter constraints on the 
statistics of young, massive stars throughout the region studied here
and to confirm or reject the hypothesis that a single power-law can be
applied to their surface density from 0 to 2.5\,pc.

As concerns the cluster-infall scenario, our results do not 
significantly alter the currently accepted picture because
the fundamental observation stays valid that the young, massive stars
are strongly concentrated toward Sgr\,A*. Nevertheless, we cannot
  yet reliably exclude the possibility that the candidates for young,
  massive stars at $R\gtrsim1$\,pc may have been deposited at their
  current location by the dissolution of a stellar cluster at larger
  distances, in an event that was unrelated to the stars within
  $R\sim0.5$\,pc of Sgr\,A*.
\citet{Gurkan05GC}, for example, showed that a $\sim 5 \times 10^5
M_{\sun}$ stellar cluster with an initial position of 10\,pc from the SMBH
can sink into the central a few\,pc within a lifetime of young, massive stars (a few Myr).  
Since the surface density profiles predicted from the inspiralling cluster scenario
strongly depend on models employed in simulations
\citep[see, e.g., simulations by][]{Kim03,Berukoff06,Fujii10GC},
it is difficult to exclude the cluster scenario 
by comparing the surface density profiles.
Kinematic information of the candidates is presumably crucial to 
discriminate the above scenarios.

We can also consider the possibility that young, massive stars at
$R \gtrsim 1$\,pc were formed in a different event (or events) 
than the ones at $R \lesssim 0.5$\,pc. 
The required density of molecular clouds against tidal destruction 
by the tidal shear from Sgr\,A* is reduced
to a few $10^{6}$\,$n_{\rm H2}$\,cm$^{-3}$ at $r=2$\,pc. 
Some clumps of molecular gas in the circumnuclear disk (CND) may in fact 
reach the necessary densities \citep{Christopher05CND, Montero-Castano09CND} 
and some observations may indicate the onset of star formation 
in parts of the CND \citep{YusefZadeh08CND}. 
It may thus be possible that the young, massive candidates 
formed at a distances from Sgr\,A* that were close to the
currently observed ones. 
Recent surveys find in fact young, massive stars 
throughout the central tens of parsecs of the MW, 
indicating that massive stars may continually form 
throughout the central region \citep[e.g.,][]{Mauerhan10IRW,Wang10HST}.

Currently, we do not have enough observational data 
to discriminate between the above scenarios. 
The direction of future work is clear. 
The immediate next step is to confirm (or discard) our candidates as
early-type stars and determine their exact properties 
through spectroscopic observations. 
This will provide not only firm evidence for the existence of 
the young, massive stars outside the central 0.5\,pc, 
but also mass and age information to understand their formation mechanism 
and their possible relation with the young stars at $R<0.5$\,pc. 
Deeper imaging and spectroscopic observations will then
be needed to search for fainter young, massive stars 
and establish their surface density as a function of location. 
Finally, proper motion measurements can provide further clues 
as to the nature and origin of the young, massive stars 
in the NSC.


\section{Conclusions}
\label{sec:Conclusion}

We have carried out a systematic search for young, massive stars
  within a projected distance of $R=2.5$\,pc from Sgr\,A*, by means of
  near-infrared narrow-band imaging with VLT/ISAAC. Our method makes use of the
  absence of the CO-bandhead absorption feature, which is ubiquitous
  in late-type stars. It is particularly prominent in the red giants
  that almost completely dominate the brightness range of our stellar
  sample, which is limited to the magnitude range of $9.75 < [2.25] <
  12.25$. We have found 63 early-type candidates at projected
  distances $R \lesssim 2.5$\,pc from the Milky Way's central black
  hole. Through comparison with previous spectroscopic studies, which
  were mainly limited to $R \lesssim 0.5$\,pc and via statistical
  considerations, based on the uncertainties of the measured stellar
  colors, we estimate the percentage of erroneously identified stars
  as low as $\sim$20\%. About half of the candidates found in this
  study have not been previously identified. All of those new
  candidates for young, massive stars are located at $R>0.5$\,pc.  Our
  study thus shows that young, massive stars can be found throughout
  the nuclear star cluster.  Follow-up studies, both spectroscopically
and via AO assisted narrow-band imaging, will be needed to 
confirm in detail and deepen the findings of this work. 
So far, we lack a consistent model of star formation 
at the Galactic Center that can explain  the presence
of young, massive stars throughout the nuclear star cluster.

\begin{acknowledgements}
  We thank the referee, D. Q. Wang, for his helpful comments.
  SN is financially supported by the Japan Society for the Promotion
  of Science (JSPS) through the JSPS Research Fellowship for Young
  Scientists.  This work was supported by KAKENHI, Grant-in-Aid for
  JSPS Fellows 20$\cdot$868, Grant-in-Aid for Research Activity
  Start-up 23840044, Grant-in-Aid for Specially Promoted Research 22000005,
  Excellent Young Researcher Overseas Visit Program, and
  Institutional Program for Young Researcher Overseas Visits.  RS
  acknowledges support by the Ram\'on y Cajal programme, by grants
  AYA2010-17631 and AYA2009-13036 of the Spanish Ministry of Science
  and Innovation, and by grant P08-TIC-4075 of the Junta de Andaluc\'ia.  
  This material is partly based upon work supported in
  part by the National Science Foundation Grant No. 1066293 and the
  hospitality of the Aspen Center for Physics.
\end{acknowledgements}


\bibliography{/Users/shogo/Documents/mypapers/Tex/references/papers}

\clearpage 
\onecolumn
\small
\begin{longtable}{cccccccccc}
\caption{Early-type star candidates. \label{Tab:EarlyCnad}} \\
\hline\hline
ID & RA\tablefootmark{a} & Dec\tablefootmark{a} & $R_{\mathrm{Sgr A*}}$\tablefootmark{b} & 
$[1.71]-[2.25]$\tablefootmark{c} & $[2.25]-[2.34]$\tablefootmark{d} & 
Name\tablefootmark{e} & Type\tablefootmark{f} & Counterpart\tablefootmark{g} &\\
 & (J2000.0) & (J2000.0) & [\arcsec] & & & & &\\
\hline
\endfirsthead
\caption{continued.}\\
\hline\hline
ID & RA\tablefootmark{a} & Dec\tablefootmark{a} & $R_{\mathrm{Sgr A*}}$\tablefootmark{b} & 
$[1.71]-[2.25]$\tablefootmark{c} & $[2.25]-[2.34]$\tablefootmark{d} & 
Name\tablefootmark{e} & Type\tablefootmark{f} & Counterpart\tablefootmark{g} &\\
 & (J2000.0) & (J2000.0) & [\arcsec] & & & & &\\
\hline
\endhead
\hline
\endfoot
 1 & 17:45:40.047 & -29:00:26.93 &  1.17 &  1.94 &  0.09 &   IRS16NW &     Ofpe/WN9 & Bu012(E1)\\
 2 & 17:45:40.122 & -29:00:27.59 &  1.19 &  2.04 &  0.08 &    IRS16C &     Ofpe/WN9 & Bu007(E1), Do129\\
 3 & 17:45:40.124 & -29:00:29.09 &  1.48 &  2.21 &  0.13 &   IRS16SW &     Ofpe/WN9 & Bu015(N)\\
 4 & 17:45:40.192 & -29:00:27.56 &  2.07 &  1.92 &  0.04 &   IRS16CC &   O9.5-B0.5I & Bu028(E1)\\
 5 & 17:45:40.039 & -29:00:30.36 &  2.26 &  2.00 &  0.05 &    IRS33N &      B0.5-1I & Bu058(E1)\\
 6 & 17:45:40.142 & -29:00:30.00 &  2.32 &  1.96 &  0.01 & E34/S2-17 &      B0.5-1I & Bu027(E1)\\
 7 & 17:45:40.094 & -29:00:31.26 &  3.24 &  2.02 &  0.16 &   IRS33SE &     Ofpe/WN9 & Bu019(E1)\\
 8 & 17:45:40.165 & -29:00:31.05 &  3.38 &  2.14 &  0.15 & E47/S3-30 &        B0-3I & Bu219(E1)\\
 9 & 17:45:40.220 & -29:00:30.84 &  3.62 &  3.74 &  0.36 &     IRS21 &          DES & Bu219(E3)\\
10 & 17:45:39.864 & -29:00:24.33 &  4.42 &   --- &  0.45 &      IRS3 &          DES & Bu581(R) \\
11 & 17:45:39.649 & -29:00:27.33 &  5.19 &  3.42 &  0.26 &     IRS6E &          DES & Bu068(R) \\
12 & 17:45:40.455 & -29:00:25.13 &  6.20 &  1.95 &  0.17 &      ---  & early$^{\mathrm{(Ba)}}$ & Bu123(E1)\\
13 & 17:45:39.958 & -29:00:20.53 &  7.65 &  1.96 &  0.12 &       E73 &        O9-BI & Bu085(E1)\\
14 & 17:45:40.324 & -29:00:20.95 &  8.06 &  2.16 &  0.10 &      ---  &          --- & Bu169(L2)\\
15 & 17:45:40.535 & -29:00:23.00 &  8.26 &  3.20 &  0.32 &    IRS10W & DES,early$^{\mathrm{(Ta)}}$ &    ---\\
16 & 17:45:40.343 & -29:00:35.44 &  8.35 &  2.29 &  0.15 &      ---  & non-early$^{\mathrm{(Ba)}}$ & Bu202(L2)\\
17 & 17:45:40.039 & -29:00:19.61 &  8.49 &  2.00 &  0.11 &       E75 &        O9-BI & Bu049(E1)\\
18 & 17:45:40.098 & -29:00:37.43 &  9.36 &  2.02 &  0.10 &      ---  & non-early$^{\mathrm{(Ba)}}$ & Bu079(E1)\\
19 & 17:45:39.324 & -29:00:28.63 &  9.41 &  2.93 &  0.13 &      ---  & late$^{\mathrm{(Bl)}}$ & Bl103, Bu183(L1)\\
20 & 17:45:39.546 & -29:00:35.01 &  9.47 &  1.93 &  0.16 &       E79 &     Ofpe/WN9 & Bu044(E1), Do117\\
21 & 17:45:39.367 & -29:00:31.57 &  9.49 &  2.06 &  0.11 &      ---  & early$^{\mathrm{(Ba)}}$ & Bu157(E1)\\
22 & 17:45:39.922 & -29:00:18.14 & 10.08 &  1.78 &  0.24 &   IRS15SW &      WN8/WC9 & Bu124(E1), Do124\\
23 & 17:45:40.047 & -29:00:17.92 & 10.18 &  1.78 &  0.07 &       E84 &        O9-BI & Bu051(E1)\\
24 & 17:45:39.363 & -29:00:33.85 & 10.58 &  1.97 &  0.10 &      ---  & early$^{\mathrm{(Ba)}}$ & Bu042(E1)\\
25 & 17:45:39.506 & -29:00:36.04 & 10.59 &  1.92 &  0.11 &      ---  &          --- & Bu096(E1)\\
26 & 17:45:40.144 & -29:00:16.50 & 11.68 &  1.90 &  0.10 &   IRS15NE &        WN8/9 & Bu102(E1), Do123\\
27 & 17:45:39.469 & -29:00:18.61 & 12.09 &  1.77 &  0.03 &      ---  & non-early$^{\mathrm{(Ba)}}$ & Bu133(L1)\\
28 & 17:45:39.375 & -29:00:37.21 & 12.61 &  2.16 &  0.07 &      ---  &          --- & Bu172(L1)\\
29 & 17:45:40.696 & -29:00:18.36 & 13.00 &  3.74 &  0.32 &      ---  &          --- &    ---\\
30 & 17:45:40.893 & -29:00:38.12 & 15.02 &  2.44 &  0.06 &      ---  &          --- & Bu185(L1)\\
31 & 17:45:39.616 & -29:00:44.46 & 17.28 &  1.74 & -0.02 &      ---  &          --- &    ---\\
32 & 17:45:40.867 & -29:00:41.73 & 17.42 &  3.82 &  0.04 &      ---  &          --- & Bu480(R)\\
33 & 17:45:39.339 & -29:00:09.22 & 21.00 &  2.13 &  0.04 &      ---  &          --- &    ---\\
34 & 17:45:39.475 & -29:00:08.12 & 21.31 &  2.32 &  0.13 &      ---  &          --- &    ---\\
35 & 17:45:41.114 & -29:00:49.62 & 25.72 &  2.68 &  0.02 &      ---  &          --- &    ---\\
36 & 17:45:42.052 & -29:00:20.02 & 27.60 &  2.72 &  0.01 &      ---  &          --- &    ---\\
37 & 17:45:38.906 & -29:00:03.50 & 28.75 &  2.09 &  0.10 &      ---  &          --- &    ---\\
38 & 17:45:39.603 & -29:00:58.80 & 31.23 &  1.96 &  0.08 &      ---  &          --- &    ---\\
39 & 17:45:39.880 & -28:59:56.49 & 31.68 &  2.03 &  0.05 &      ---  &          --- &    ---\\
40 & 17:45:39.416 & -29:00:58.78 & 31.75 &  3.05 & -0.03 &      ---  &          --- &    ---\\
41 & 17:45:41.117 & -28:59:57.27 & 33.91 &  2.23 &  0.04 &      ---  &          --- &    ---\\
42 & 17:45:38.668 & -28:59:57.79 & 35.25 &  1.92 &  0.13 &      ---  &          --- &    ---\\
43 & 17:45:42.266 & -29:00:49.47 & 36.19 &  2.53 &  0.01 &      ---  &          --- &    ---\\
44 & 17:45:37.524 & -29:00:46.02 & 37.56 &  3.84 &  0.01 &      ---  & late$^{\mathrm{(Bl)}}$ &    Bl84\\
45 & 17:45:37.436 & -29:00:08.73 & 39.27 &  2.14 &  0.10 &      ---  &          --- &    ---\\
46 & 17:45:37.122 & -29:00:38.29 & 39.61 &  2.27 &  0.05 &      ---  &          --- &    ---\\
47 & 17:45:36.878 & -29:00:33.15 & 41.79 &  2.15 &  0.03 &      ---  &          --- &    ---\\
48 & 17:45:40.478 & -28:59:42.85 & 45.61 &  2.91 &  0.03 &      ---  &          --- &    ---\\
49 & 17:45:41.686 & -29:01:08.96 & 46.21 &  2.59 &  0.08 &      ---  &          --- &    ---\\
50 & 17:45:36.561 & -29:00:12.07 & 48.37 &  2.13 &  0.10 &      ---  &          --- &    ---\\
51 & 17:45:37.616 & -28:59:47.42 & 51.63 &  2.49 &  0.00 &      ---  &          --- &    ---\\
52 & 17:45:35.917 & -29:00:17.02 & 55.21 &  2.14 &  0.01 &      ---  &          --- &    ---\\
53 & 17:45:39.082 & -28:59:33.00 & 56.52 &  1.07 &  0.03 &       fg  &          --- &    ---\\
54 & 17:45:43.872 & -29:00:01.32 & 56.96 &  2.27 &  0.05 &      ---  &          --- &    ---\\
55 & 17:45:40.513 & -28:59:31.05 & 57.39 &  2.65 &  0.05 &      ---  &          --- &    ---\\
56 & 17:45:35.623 & -29:00:11.96 & 60.15 &  2.32 &  0.08 &      ---  &          --- &    ---\\
57 & 17:45:37.013 & -29:01:24.71 & 69.15 &  2.25 &  0.06 &      ---  &          --- &    ---\\
58 & 17:45:38.583 & -29:01:36.43 & 70.95 &  2.20 &  0.08 &      ---  &          --- &    ---\\
59 & 17:45:41.512 & -28:59:18.46 & 72.27 &  2.56 & -0.01 &      ---  &          --- &    ---\\
60 & 17:45:41.819 & -28:59:18.38 & 73.52 &  2.29 & -0.00 &      ---  &          --- &    ---\\
61 & 17:45:44.737 & -29:01:09.06 & 73.99 &  5.33 &  0.33 &      ---  &          --- &    ---\\
62 & 17:45:43.062 & -28:59:24.95 & 74.56 &  2.52 &  0.01 &      ---  &          --- &    ---\\
63 & 17:45:35.700 & -29:01:23.08 & 79.14 &  2.02 &  0.02 &      ---  &          --- &    ---\\
\end{longtable}
\tablefoot{ \tablefoottext{a}{Typical positional uncertainty is
    $0\farcs1 - 0\farcs3$ (see \S \ref{subsec:Phot}).}
  \tablefoottext{b}{Distance from Sgr A*.}
  \tablefoottext{c}{Magnitude difference between $[1.71]$ and
    $[2.25]$, after the color correction (\S
    \ref{subsec:correction}).}  \tablefoottext{d}{Magnitude difference
    between $[2.25]$ and $[2.34]$, after the color correction (\S
    \ref{subsec:correction}) and the relative extinction correction
    (\S \ref{subsec:171CMD}).}  \tablefoottext{e}{Name(s) of
    counterparts in \citet{Paumard06} and \citep{Lu09}.  The source \#
    53 is probably a foreground star, because of its very blue color
    in $[1.71] - [2.25]$. } \tablefoottext{f}{Spectral type from
    \citet{Paumard06}, while the superscript '(Ba)' indicates the
    spectral type (early or non-early) from \citet{Bartko10}, 
    '(Ta)' from \citet{Tanner05DES}, and
    '(Bl)' from \citet{Blum03}.}  \tablefoottext{g}{Counterparts from
    \citet[][superscript 'Bl']{Blum03}, 
    \citet[][superscript 'Bu']{Buchholz09} and
    \citet[][superscript 'Do']{Dong11HST}, and their source ID.  
    E1, E3, L1, L2, N, and R
    denote quality 1 early-type candidates, quality 3 early-type
    candidates, quality 1 late-type candidates, quality 3 late-type
    candidates, sources with noisy SED, and strongly reddened sources,
    respectively, in \citet{Buchholz09}.}}

\newpage

\end{document}